\pdfoutput=1

\documentclass{article}
\usepackage{natbib,bm}
\bibpunct[; ]{(}{)}{,}{a}{}{;}
\usepackage[super]{nth}
\usepackage[utf8]{inputenc}
\usepackage{gensymb}
\usepackage{hyperref}
\usepackage{graphicx} 
\usepackage{subcaption} 
\usepackage{amsmath, amssymb, bm}
\usepackage{units} 
\usepackage[a4paper]{geometry}
\usepackage[labelfont=bf]{caption}
\usepackage{xcolor} 
\usepackage{float} 
\usepackage{authblk} 

\usepackage[flushmargin]{footmisc}

\definecolor{thr0}{RGB}{128,0,38}
\definecolor{thr1}{RGB}{227,26,28}
\definecolor{thr2}{RGB}{254,178,76}
\newcommand{\actlegendbay}{
\setlength{\fboxsep}{5pt} 
\setlength{\fboxrule}{1pt} 
\begin{picture}(130,5)
\put(5,10){\small 0\%}
\put(22,13){\fcolorbox{black}{thr0}{\null}}
\put(45,10){\small 1\%}
\put(62,13){\fcolorbox{black}{thr1}{\null}}
\put(85,10){\small 2\%}
\put(102,13){\fcolorbox{black}{thr2}{\null}}
\end{picture}}

\definecolor{corr1}{RGB}{50,50,50}
\definecolor{corr2}{RGB}{169,169,169}
\newcommand{\actlegendclass}{
\setlength{\fboxsep}{5pt} 
\setlength{\fboxrule}{1pt} 
\begin{picture}(130,5)
\put(10,10){\small FWER}
\put(42,13){\fcolorbox{black}{corr1}{\null}}
\put(65,10){\small FDR}
\put(90,13){\fcolorbox{black}{corr2}{\null}}\end{picture}}

\def\bfC{\mathbf C}

\def\bfG{\mathbf G}

\def\bfI{\mathbf I}

\def\bfQ{\mathbf Q}

\def\bfX{\mathbf X}

\def\bfx{\mathbf x}
\def\bfy{\mathbf y}

\def\bfbeta   {\bm \beta}

\def\bfepsilon{\bm \epsilon}

\def\bftheta  {\bm \theta}

\newcommand{\bfzero}{{\mathbf 0}}

\def\boldfacefake#1{\kern-4pt
    \hbox{ \mathsurround=0pt
    \hbox to 0.4pt{$#1$\hss}\hbox to 0.4pt{$#1$\hss}\hbox {$#1$}}}

\newcommand{\be}{\begin{eqnarray}}
\newcommand{\ee}{\end{eqnarray}}
\newcommand{\ba}{\begin{eqnarray*}}
\newcommand{\ea}{\end{eqnarray*}}
\newcommand{\bc}{\begin{center}}
\newcommand{\ec}{\end{center}}
\newcommand{\btab}[1]{\begin{tabular}{#1}}
\newcommand{\etab}{\end{tabular}}

\title{Longitudinal surface-based spatial Bayesian GLM reveals complex trajectories of motor neurodegeneration in ALS}
\author[1]{Amanda F. Mejia\thanks{Corresponding author: Amanda Mejia, mandy.mejia@gmail.com}}
\affil[1]{Department of Statistics, Indiana University, Bloomington, IN, USA}
\author[2]{Vincent Koppelmans}
\affil[2]{Department of Psychiatry, University of Utah, Salt Lake City, UT, USA}
\author[3]{Laura Jelsone-Swain}
\affil[3]{Department of Psychology, University of South Carolina Aiken, Aiken, SC, USA}
\author[4]{Sanjay Kalra}
\affil[4]{Division of Neurology, Department of Medicine, University of Alberta, Edmonton, AB, Canada}
\author[2]{Robert C. Welsh}

\date{}

\begin{document}

\maketitle

\begin{abstract}
Longitudinal fMRI datasets hold great promise for the study of neurodegenerative diseases, development and aging, but realizing their full potential depends on extracting accurate fMRI-based measures of brain function and organization in individual subjects over time. This is especially true for studies of rare, heterogeneous and/or rapidly progressing neurodegenerative diseases, which often involve a small number of subjects whose functional features may vary dramatically both across subjects and over time, making traditional group-difference analyses of limited utility. One such disease is amyotrophic lateral sclerosis (ALS), a severe disease resulting in extreme loss of motor function and eventual death. Here, we extend our advanced individualized statistical approach to analyze a rich longitudinal dataset containing 190 hand clench task fMRI scans from 16 ALS patients (78 scans) and 22 age-matched healthy controls (112 scans). Specifically, we adopt our cortical surface-based spatial Bayesian general linear model (GLM), which has high power and precision to detect activations in individual subjects, and we propose a novel longitudinal extension to leverage information shared across visits. We perform all analyses in participant space in order to better preserve anatomical and functional features of each individual, an approach facilitated by the high power of spatial Bayesian modeling. Using a series of longitudinal mixed-effects models to subsequently study the relationship between task activation and ALS disease progression, we observe for the first time an inverted U-shaped trajectory of motor activations: at relatively mild motor disability we observe enlarging activations, while at higher levels of motor disability we observe severely diminished activation, reflecting progression toward complete loss of motor function. Furthermore, we observe distinct trajectories depending on clinical progression rate, with faster progressors exhibiting more extreme hyper-activation and subsequent hypo-activation occurring at an earlier stage of disability. These differential trajectories suggest that initial hyper-activation is likely attributable to loss of inhibitory neurons, part of a more general process of motor neuron loss in ALS.  Earlier studies employing more limited sampling designs and using traditional group-difference analysis approaches were only able to observe the initial hyper-activation, which was assumed to be due to a compensatory process. Our more nuanced findings substantially advance scientific understanding of the ALS disease process. This study provides the first real-world example of how this advanced surface-based spatial Bayesian modeling approach furthers scientific understanding of neurodegenerative disease, particularly in a longitudinal context where it is critical to obtain robust and reliable individual measures of brain function and organization. This approach also holds promise for the study of other time-varying processes such as development and aging. The surface-based spatial Bayesian GLM, including our new longitudinal extension, is implemented in a user-friendly R package. \\


\end{abstract}

\section{Introduction}


Longitudinal fMRI studies are a powerful tool for examining functional brain changes occurring within individuals in the context of neurodegenerative diseases, development and normal aging \citep{telzer2018methodological}. Longitudinal studies can account for measures reflective of heterogeneity between and across participants over time (e.g., disease burden) 
\citep{Lawrence:2017ig, Kassubek:2014ek} and are key for the development of biomarkers \citep{Turner:2010fd}, an area of ongoing interest and development in neuroimaging research \citep{woo2017building}.  

To utilize longitudinal fMRI datasets to their full potential, it is critical to develop and employ statistical methods for accurate individual-level analysis, rather than traditional group-average and group-difference analysis.  This is particularly true for longitudinal studies of rare, heterogeneous and/or rapidly progressing neurodegenerative diseases---which are often difficult and expensive to acquire---as they typically involve small numbers of participants whose functional brain features may vary markedly across subjects and over time.  Individualized analytical methods would facilitate the use of such studies to understand the dynamics of neurodegeneration, to develop neuroimaging biomarkers, and to ultimately translate research findings to monitor disease progression and evaluate treatment efficacy clinically.  Individualized analyses are also less likely than group-average analyses to require spatially warping participants' brains to a standard template.  Such warping or normalization can induce errors and inaccuracies, particularly in individuals with neurodegenerative disease and even in normal aging \citep{Eloyan:2014hz, Kolinger:2021bc}. 

Individual-level estimates of activation based on conventional task fMRI analysis methods have unfortunately been found to exhibit poor reliability \citep{elliott2020test}, in part due to sub-optimal statistical approaches \citep{monti2011statistical}.  By far the most popular method for task fMRI analysis is the classical general linear model (GLM) \citep{Friston:1995vs}. In this massive univariate approach, at the first level a separate linear model is fit at every voxel relating the observed BOLD activity to the expected response to each task or stimulus. At the second level, participant-level estimates are pooled to produce group average estimates of activation or differences between groups or conditions.  Historically, the classical GLM has been considered sufficient in group-average fMRI studies, where the focus is on estimation of robust effects that are common across most participants  \citep{mumford2009simple}. However, effects that are unique to individuals or states are likely to be washed out in a group analysis \citep{Stern:2009ex, Gupta:2010ff}, and the classical GLM has been shown to have low estimation efficiency and power for subject-level analysis \citep{mejia2020bayesian}.  Additionally, many task fMRI analyses are performed in volumetric space, which has a number of drawbacks, including smoothing across tissue classes and distal cortical areas representing distinct functional regions \citep{brodoehl2020surface}.

Here, we adopt and extend our advanced individualized fMRI analysis approach, the cortical surface-based spatial Bayesian GLM \citep{mejia2020bayesian}, to better understand the longitudinal disease process in amyotrophic lateral sclerosis (ALS), a rare, rapidly progressing and heteroegenous neurodegenerative disease. This approach avoids data smoothing, leverages spatial dependencies along the cortical surface, and avoids the need for multiplicity correction---and the resulting loss of power---in identifying areas of activation. It yields substantially more accurate and powerful activations in individual subjects \citep{mejia2020bayesian}.  We propose a novel longitudinal model extension, which leverages information shared across multiple visits. Our analyses are performed entirely in native space to preserve the unique anatomical and functional features of each participant \citep{Gray:2012em, Kolinger:2021bc}, using the size of activations above a certain effect size as a summary measure that can be used in subsequent longitudinal analysis. 

Specifically, we analyze a rich longitudinal study including $78$ motor task fMRI scans from $16$ ALS participants and $112$ scans from $22$ age-matched healthy control (HC) participants. Most ALS participants were enrolled soon following clinical diagnosis, and follow-up scans occurred roughly every $3$ months for as long as participants were willing and able to participate in the study. This study is unique in terms of the high frequency and long duration of sampling, providing an opportunity to understand how disease progression evolves dynamically in different individuals.  We apply our longitudinal surface-based spatial Bayesian GLM to each participant’s data and extract summary measures of activation at each visit. We then analyze these summary measures using a series of longitudinal mixed effects models to study the relationship between brain activation and functional disability in ALS over time. 

Previous studies have documented ALS-related motor disruptions using a variety of measures, including task activation  \citep{Konrad:2002hl,Schoenfeld:2005eh,Konrad:2006hy,Poujois:2013hn,Stoppel:2014bj}, functional connectivity \citep{JelsoneSwain:2010gj,Menke:2018ed}, cortical thickness \citep{Verstraete:2010gr,Agosta:2012gx}, and structural connectivity \citep{Douaud:2011hb,Menke:2012dc,Chapman:2014cg,Muller:2016er}.  Task activation studies specifically have observed hyper-activation in patients with ALS during motor tasks \citep{Konrad:2002hl, Schoenfeld:2005eh, Poujois:2013hn}. However, since these studies have been based on more limited sampling designs \citep{Menke:2018ed} and employed group-average analysis approaches, they provide limited insight into the heterogeneous and typically rapid disease process characterizing ALS. 

Our longitudinal analyses reveal an inverted U-shaped trajectory of motor activation associated with disease progression in ALS, with hyper-activation initially occurring at relatively low levels of motor disability, followed by dramatic loss of activation occurring at with more severe motor disability. To our knowledge, this is the first time this inverted U-shaped trajectory has been observed in ALS. Further, we observe systematic differences by clinical progression rate, with fast progressors exhibiting more extreme hyper-activation and more severe loss of activation earlier in the disease process.  
Despite the long-standing interest in spatial Bayesian techniques for task fMRI analysis \citep{friston2003posterior, woolrich2009bayesian}, to our knowledge this is the first study of neurodegenerative disease that has employed a surface-based spatial Bayesian GLM, and illustrates the power of this approach to extract new scientific insights, especially when applied to rich longitudinal fMRI studies.



\section{Materials and Methods}
\label{sec:methods}

\subsection{Participants}

ALS participants were recruited through the University of Michigan ALS clinic and had been diagnosed as having probable or definite ALS per the El Escorial Criteria \citep{Brooks:1994wd}.  All ALS participants had limb onset rather than bulbar onset disease. The project was approved by the University of Michigan Institutional Review Board (HUM00000219). All participants gave written informed consent before participation. Healthy controls were recruited from the local community through community advertising. The healthy controls were balanced for sex and age. Participants underwent magnetic resonance imaging (MRI) at multiple research visits. For convenience for the ALS participants, research visits were scheduled to coincide with clinical visits.  Visits were scheduled at roughly 3-month intervals while minimizing travel burden to participants with ALS. Participants with ALS continued into the study until they could no longer tolerate the MRI session, decided on their own accord to no longer participate, or passed away. The sample used for analysis included 190 scanning visits from 16 ALS participants (78 visits) and 22 HC participants (112 visits). The number of visits per participant ranged from 3 to 10 (median = 4.5, mean = 5). Participants’ visit timing is shown in Supplementary Fig.\ \ref{fig:visit_timing}.

\begin{figure}[t]
    \centering
    \includegraphics[page=1]{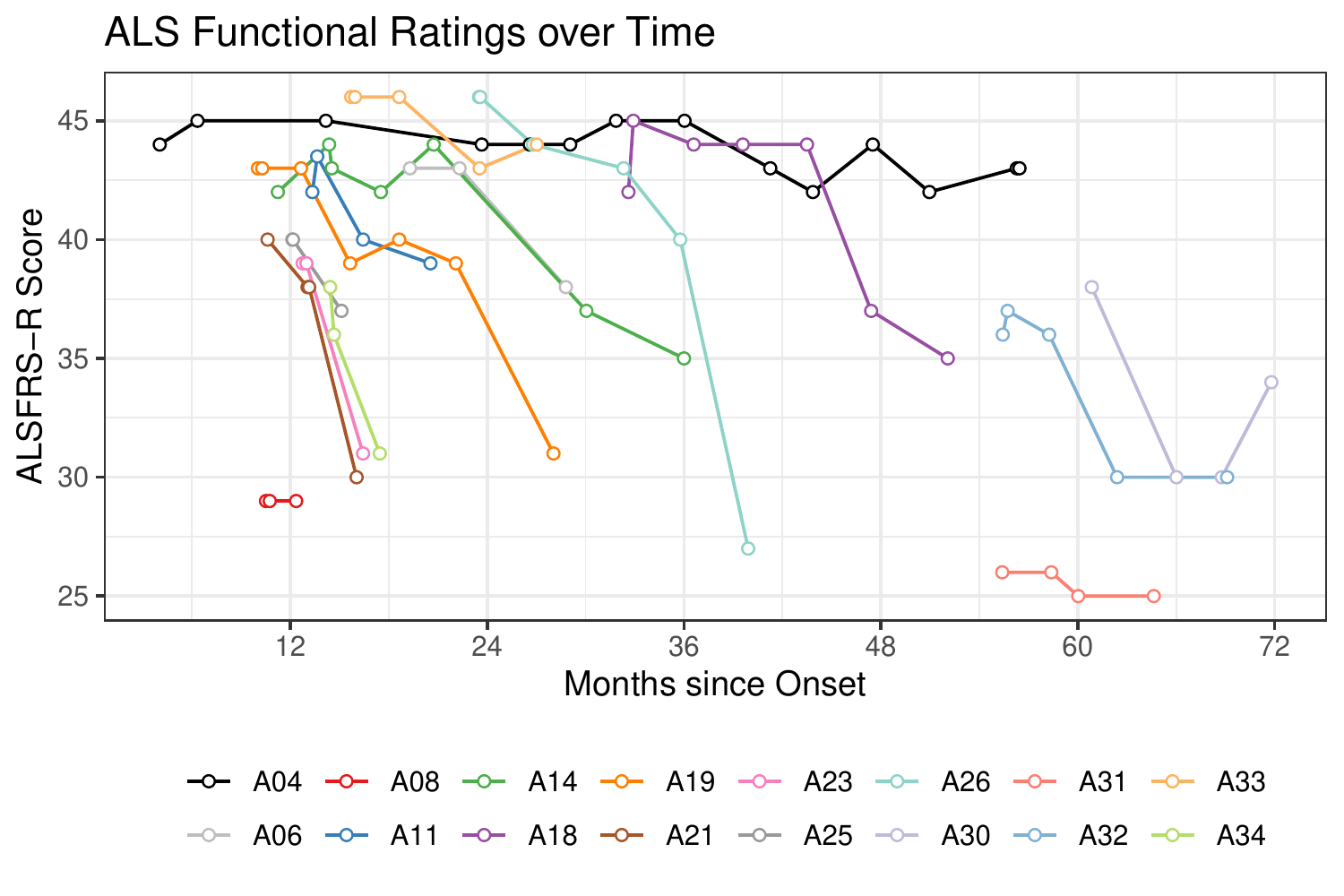}
    \caption{ \small \textbf{ALSFRS-R disease trajectories over time.} ALS Functional Rating Scale (revised) (ALSFRS-R) scores range from 48 (no impairment) to 0 (total impairment). Each line represents an individual participant with ALS. Note that for many participants, the first two visits occurred in quick succession and appear overlapping on the plot. These clinical trajectories illustrate the wide heterogeneity of the disease process, with some patients exhibiting rapid clinical progression (e.g., A08, A21, A23, A25, A34) while others exhibit very slow progression (e.g., A04).}
    \label{fig:ALSFRS}
\end{figure}

At the time of each scanning visit, ALS participants had function assessed with the ALS Functional Rating Scale (revised) (ALSFRS-R), a standard instrument that is widely used in clinical care and clinical research \citep{Rooney:2017hk}. ALSFRS-R scores range from 48 (no impairment) to 0 (total impairment).  Fig.\ \ref{fig:ALSFRS} displays ALSFRS-R trajectories for each ALS participant in the study. All but two participants had ALSFRS-R scores over 35 at the first visit, indicating a relatively mild disease state.  Fig.\ \ref{fig:ALSFRS} shows that many ALS participants experienced wide range of disability levels through the course of the study. More details on each individual with ALS is provided in Supplementary Section \ref{app:participants}.

\subsection{MRI Data Collection}

 MRI data were collected on a GE 3T Excite 2 scanner (General Electric, Milwaukee, Wisconsin). A high-resolution $T_1$-weighted image was acquired (3D SPGR, IR 500ms, $15^{\degree}$ flip angle, TR = 9.036ms, TE = 1.84ms, $256\times 256\times 160$ matrix, $1.102\times 1.102\times 1.2$mm resolution). For image coregistration, we also collected a lower resolution $T_1$-weighted image ($85^{\degree}$ flip angle, TR = 250ms, TE = 5.70ms, $256\times160$ matrix, 3mm slice thickness and no skip) at the same spatial locations as the functional time-series data. Blood oxygenation level dependent (BOLD) data were collected using a reverse-spiral k-space trajectory sequence \citep{Noll2002RapidMI}, and reconstructed off-line using a gradient descent algorithm \citep{Noll:2004cz}. 
 
 BOLD fMRI data were collected for 4 tasks: right hand finger tapping, right first clenching, left hand finger tapping, and left fist clenching. All tasks used a paced-block design, in which a visual target appeared every 1.5 seconds for the participant to execute the given task. This visual cue appeared 20 times in each block, followed by 15 seconds of a fixation cross-hair. A block design was used for robustness and less sensitivity to variation in the hemodynamic response across individuals \citep{Liu:2001bo,Shan:2013cf}. Task difficulty was matched on hand-strength. Just prior to the scan, participants had their hand strength measured 3 times with a dynanometer (Jamar Hydraulic Hand Dynanometer, Model SD081028935). In the scanner, participants squeezed a resistive hand exerciser (Sammons-Preston Model 56573) set to approximately 10\% of their mean measured hand strength. The task was practiced outside the scanner to ensure participant compliance. The block-rest cycle was repeated for a total of 6 times. Each BOLD time-series run was limited to a single task, resulting in 4 separate runs. BOLD data were collected with a $T^*_2$-weighted gradient-echo, spiral-readout sequence ($90^{\degree}$ flip angle, TR = 2s, TE = 30ms, 64x64 matrix, 3 mm slice thickness and no skip, 220mm field-of-view, sequential and ascending acquisition, 40 slices). The first four $T^*_2$ volumes at the beginning of the time-series sequence were excited but not recorded to allow for magnetization equilibrium. We limit analyses to the right hand first clenching task, since clenching is simpler to execute than the more complex finger tapping sequence (index-to-thumb followed by middle-to-thumb) and more participants were able to complete the right hand tasks.

\subsection{Image Processing}

We constructed a custom pipeline to process the BOLD time series to the cortical surface for statistical analysis. See Supplementary Section \ref{app:processing} for details. The goal of this pipeline was to produce surface BOLD time-series data registered to a participant-specific template to allow for longitudinal modeling of co-registered visits for a given participant, while respecting participant-specific cortical anatomy for more accurate spatial dependence modeling in the Bayesian analysis described in Section \ref{sec:GLM} below.

First, the BOLD data were slice-time corrected, realigned, and registered to the N4 non-uniformity corrected $T_1$-weighted structural scan. A participant template was created based on the bias field corrected $T_1$-weighted images of all sessions using iterative rigid body registration through Advanced Normalization Tools (ANTs)\citep{Avants:2011kk}. For each participant, this template was processed through FreeSurfer \citep{Fischl:2012el} to result in a model of the pial surface and a corresponding spherical surface. No data smoothing was performed, since the spatial Bayesian model implicitly smooths task activation maps at an optimal level. 

A participant-specific sensorimotor mask consisting of four FreeSurfer sensorimotor areas was constructed to limit the location of statistical estimation. To reduce computational load of Bayesian model estimation without significant loss of spatial resolution, we resampled the pial surfaces, BOLD data, and masks to 10,000 vertices per hemisphere. The resampled mask contained approximately 1,500 vertices per hemisphere. Supplementary Fig.\ \ref{fig:mesh_ALS} displays the resampled pial surfaces and mask for each hemisphere for one individual with ALS.

We identified and removed highly noisy volumes and sessions based on a data-driven scrubbing technique \citep{mejia2017pca}, resulting in exclusion of six sessions in total (see Supplementary Section \ref{app:model} for details). Prior to model fitting, we centered and scaled the BOLD data to units of local percent signal change. We also regressed nuisance signals from the fMRI data and design matrix, including the six rigid body realignment parameters, their first derivatives, and linear and quadratic trends. 

\subsection{Statistical Analysis}\label{sec:GLM}

The block design for right hand clenching was convolved with a canonical hemodynamic response function (HRF), a double gamma-variant function \citep{Shan:2013cf}. We also included the temporal derivative of the task stimulus function to allow for differences in HRF onset timing across the brain, participants and visits (Supplementary Fig.\ \ref{fig:HRF}). No prewhitening was performed, since inspection of the residuals revealed little to no temporal dependence, likely due to the relatively long TR and inclusion of the HRF derivative.

We fit a longitudinal spatial Bayesian GLM, described next, to produce estimates of activation amplitude and areas of activation for each participant and each visit. The model was fit within each hemisphere's cortical surface separately. Areas of activation were based on the joint posterior distribution of activation amplitude, while controlling the family-wise error rate (FWER) at a significance level of $\alpha=0.05$. We computed the total size of activation by summing the surface area associated with each vertex identified as activated.

For comparison, we also fit a classical ``massive univariate'' GLM, including identifying areas of activation by performing a t-test at every location. We corrected for multiplicity within each hemisphere using Bonferroni correction to control the FWER and the Benjamini-Hochberg procedure \citep{benjamini1995controlling} to control the false discovery rate (FDR). Note that while Bonferroni correction is often considered overconservative in traditional whole brain analysis, here we are performing a much smaller number of tests (approximately 1,500 per hemisphere). A significance level of $\alpha=0.05$ was used, as in the spatial Bayesian GLM.

\subsubsection{Longitudinal spatial Bayesian modeling}

We adapted the spatial Bayesian GLM proposed by \cite{mejia2020bayesian} for analysis of task fMRI on the cortical surface. The original model was designed for single-subject, single-session analysis. We proposed a novel longitudinal extension to allow for subject- and visit-specific estimation and areas of activation, while leveraging information shared across visits. The details of this model are described in Supplementary Section \ref{app:model}. Briefly, the spatial Bayesian GLM leverages similarities in activation patterns across the cortex, resulting in smoother, more accurate estimates and areas of activation compared with a massive univariate approach \citep{mejia2020bayesian}. In addition, the Bayesian model has high power to identify areas of activation. Instead of the traditional approach of hypothesis testing followed by multiple comparisons, this model utilizes the joint posterior distribution of activation to identify a single set of locations that have high probability of being activated, achieving FWER control.

The high statistical power of the spatial Bayesian GLM may result in a phenomenon where large areas of low effect size are deemed significantly activated \citep{cremers2017relation}.  Therefore, a scientifically relevant effect size, $\gamma$, is often specified to avoid detecting irrelevant activations. For example, an effect size of $\gamma=1\%$ can be adopted to identify only locations that exhibit $>1\%$ local signal change due to the task.  An effect size of $\gamma=0\%$ would correspond to the traditional hypothesis testing framework used in the classical GLM.   Here, we consider three effect sizes: $\gamma=0\%$, $1\%$ and $2\%$.

Model fitting was performed in using the \texttt{BayesfMRI} R package (version 1.8) (\url{https://github.com/mandymejia/BayesfMRI/}). 

\subsubsection{Relating size of activation to disability}

To examine the relationship between size of motor activation and ALS disability, we fit a series of random intercept models for each effect size $\gamma$ and each hemisphere (left/right). To avoid the undue influence of temporal outliers on the regression fit, we limited ALS participant data to a 2-year window of maximal change in ALSFRS-R. This resulted in removal of the first visit for participant A14 (Fig.\ \ref{fig:ALSFRS}). Additionally, participant A04 had a very slow disease trajectory (Fig.\ \ref{fig:ALSFRS}) and many visits spanning a long duration (Fig.\ \ref{fig:visit_timing}). To avoid undue influence of this unusual individual, their data was excluded from model fitting.

In lieu of time since symptom onset, given the heterogeneity of the rate of disease progression, we constructed three predictors related to disability in ALS, adopting from \cite{Rooney:2017hk}: \textit{Total Disability} ($D^{tot}_{ij}$), \textit{Hand Motor Disability} ($D^{hand}_{ij}$) and \textit{Other Disability} ($D^{oth}_{ij}$). For participant $i$ at visit $j$,  

\begin{equation}
D^{tot}_{ij} = 1 - \frac{ALSFRS_{ij}}{48},\quad
D^{hand}_{ij} = 1 - \frac{ALSFRS_{ij}^{hand}}{12},\text{ and }
D^{oth}_{ij} = 1 - \frac{ALSFRS_{ij}^{other}}{36},
\end{equation}

where $ALSFRS_{ij}$ is the total ALSFRS-R score; $ALSFRS_{ij}^{hand}$ is the sum of three ALSFRS-R item scores for tasks related primarily to hand function: handwriting, cutting (with or without gastronomy), and dressing/hygiene; $ALSFRS_{ij}^{other}$ is the sum of the nine remaining ALSFRS components (speech, salivation, swallowing, turning in bed and adjusting bed clothes, walking, climbing stairs, dyspnea, orthopniea, and respiratory insufficiency). $D^{tot}_{ij}$, $D^{hand}_{ij}$ and $D^{oth}_{ij}$ each range from $0$ (no disability) to $1$ (total disability).  

Let $A_{ij}$ be the size of the area of activation above a given effect size for participant $i$ at visit $j$. The total disability random intercept model for the ALS group is

\begin{equation}\label{eqn:lmer_ALS_tot}
    A_{ij} = \beta_0  + b_{0i}  + f (D^{tot}_{ij}) + \epsilon_{ij},
    \quad \epsilon_{ij} \sim N(0, \sigma^2),
\end{equation}

where $\beta_0$ represents the average size of activation when \textit{Total Disability} is zero; $b_{0i}$ represents the random deviation for subject $i$; and the spline function $f(\cdot)$ allows for a non-linear relationship between \textit{Total Disability} and activation size (see Supplementary Section \ref{app:processing} for details). Note that we do not include age as a predictor to avoid conflation with the random intercept or the disability measures. The hand motor disability model for the ALS group is

\begin{equation}\label{eqn:lmer_ALS}
    A_{ij} = \beta_0  + b_{0i}  + f (D^{hand}_{ij}) + \beta_1 D^{oth}_{ij} + \epsilon_{ij},
    \quad \epsilon_{ij} \sim N(0, \sigma^2),
\end{equation}

where $\beta_0$ represents the average size of activation when \textit{Hand Motor Disability} and \textit{Other Disability} are both zero; $b_{0i}$ represents the random deviation for subject $i$; $\beta_1$ represents the average change in size of activation associated with a 1-unit increase in \textit{Other Disability}. The spline function $f(\cdot)$ allows for a non-linear relationship between \textit{Hand Motor Disability} and activation size. This model form was determined by a series of likelihood ratio tests (LRTs) (Supplementary Section \ref{app:processing}). Alternative model formulations were substantially worse in terms of predictive accuracy and Akaike information criterion (AIC) \citep{akaike1998information}. Note that these models incorporate both longitudinal data sources on participants with ALS: brain activation (fMRI) and disability (ALSFRS-R).

To investigate the role of disease progression on the relationship between size of activation in disability in ALS, we divided subjects into three groups of progressors based on their progression rate \citep{Ellis:1999bx}: fast, moderate and slow (see Table \ref{tab:progression}). Only one participant (A04) was classified as a slow progressor. We therefore limited this analysis to a comparison of fast (5 participants) and moderate (10 participants) progressors. The model in Eqn. \ref{eqn:lmer_ALS} was re-estimated within both groups.

\begin{table}
    \centering
    \begin{tabular}{lll}
    \textbf{Progression Rate} & \textbf{Progression Rate Range} & \textbf{ALS Participants} \\
    \hline
    Slow & $<0.1$ per month & A04  \\
    Moderate & $0.1$ to $0.69$ per month & All other ALS participants \\
    Fast & $\geq 0.7$ per month & A08, A21, A23, A25, A34 \\
    \hline
    \end{tabular}
    \caption{\textbf{Progression Rate Groups.} Progression rate was calculated as the average decrease in ALSFRS-R score (from the maximum value of 48, representing no disability) per month from disease onset to the last visit of each subject. The five fast progressors are those seen as exhibiting early decline in Fig. \ref{fig:ALSFRS}.}
    \label{tab:progression}
\end{table}

We fit a separate model for HC participants, as they did not complete the ALSFRS-R. To avoid undue influence of individuals with an unusually high number of visits (Fig.\ \ref{fig:visit_timing}), we excluded any visits occurring over 2 years (730 days) post-enrollment. Time in study was considered as a predictor but was found to be insignificant based on a LRT. Therefore, we adopted the following intercept-only model for HC participants:

\begin{equation}\label{eqn:lmer_HC}
    A_{ij} = \beta_0  + b_{0i}  + \epsilon_{ij},
    \quad \epsilon_{ij} \sim N(0, \sigma^2).
\end{equation}

where $\beta_0$ represents the mean size of activation across HC participants and $b_{0i}$ represents the random deviation for participant $i$.  The models in equations (\ref{eqn:lmer_ALS_tot}) to (\ref{eqn:lmer_HC}) were fit in R using the \texttt{lmer} function from the \texttt{lme4} package, version 1.1-23 \citep{lme4}.

\section{Results}
\label{sec:results}

We first performed a validation of the areas of activation produced by the longitudinal spatial Bayesian GLM, compared with the classical GLM. See Supplementary Section \ref{app:validation} for details. We observed that the Bayesian GLM produced noticeably smoother estimates and much larger areas of activation at a given effect size, while maintaining FWER control. We also analyzed the longitudinal stability of HC participants’ areas of activation, which should not change substantially over time, and found that the Bayesian GLM produced more stable results. For the subsequent analysis, we therefore adopted the Bayesian GLM.

\begin{figure}
\centering
\begin{tabular}{ccc}
& \multicolumn{2}{c}{\large \textbf{Healthy Control Participant}} \\[5pt]
& {{Activation Amplitude}} & {{Area of Activation}} \\[5pt]
\begin{picture}(0,60)\put(-5,30){\rotatebox[origin=c]{90}{C07 visit 1}}\end{picture} & 
\fbox{\includegraphics[width=2.8in, trim=0 16cm 0 3.5cm, clip]{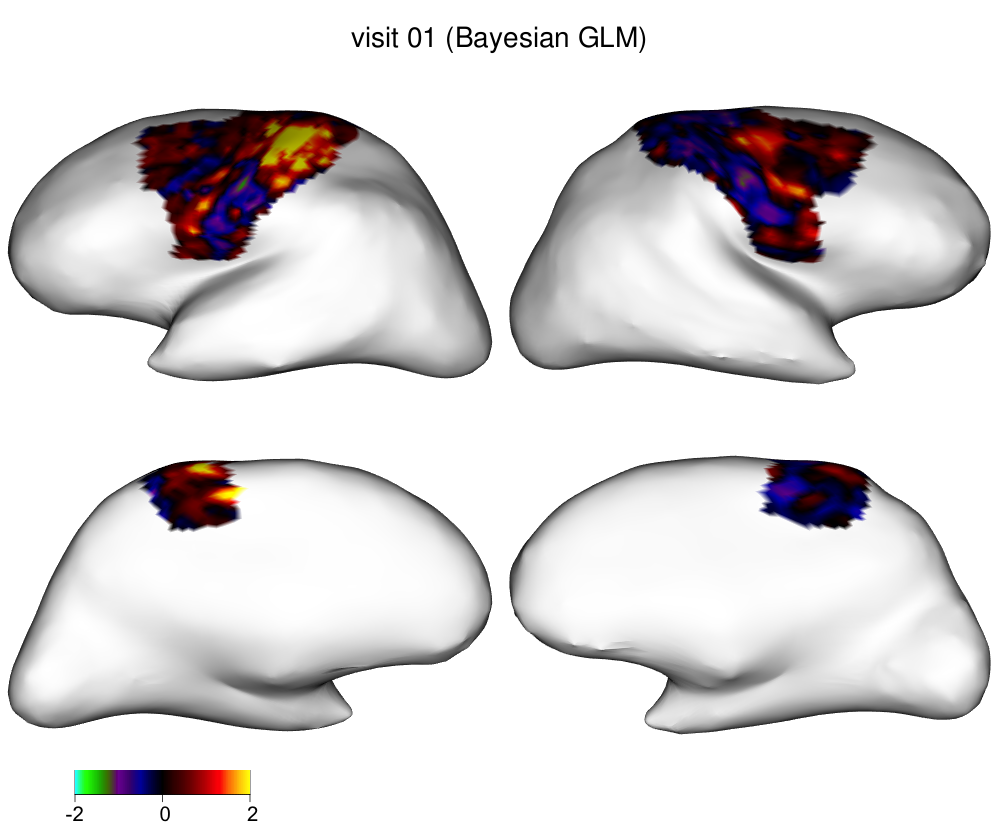}} &
\fbox{\includegraphics[width=2.8in, trim=0 13.5cm 0 3.5cm, clip]{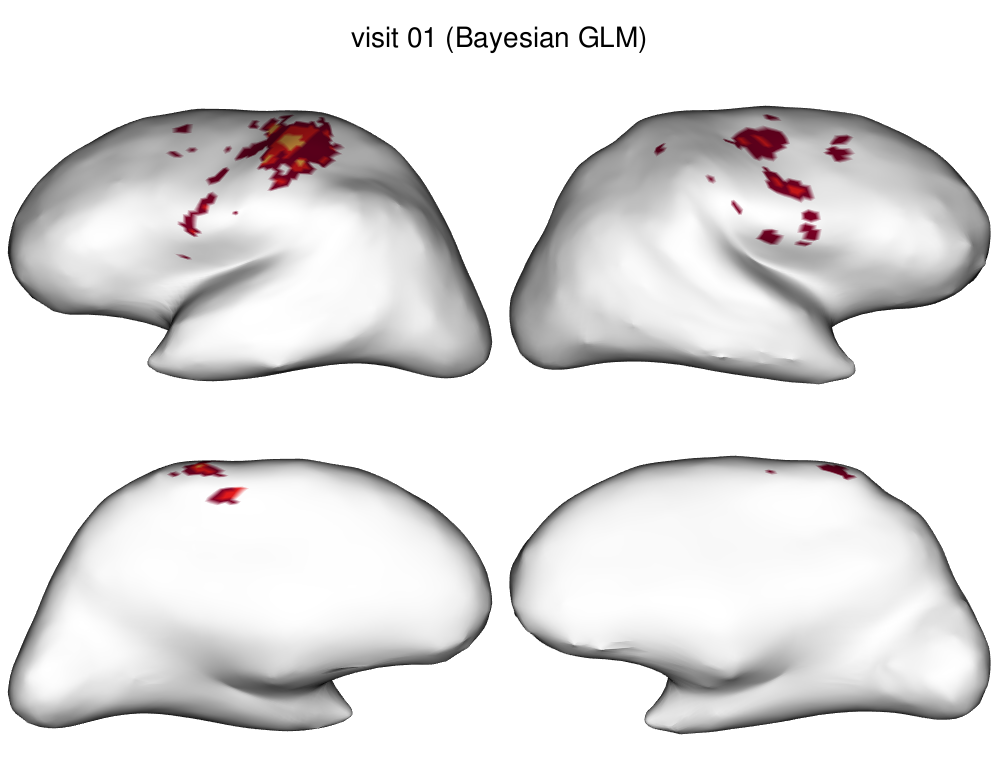}} \\[10pt]
\begin{picture}(0,60)\put(-5,30){\rotatebox[origin=c]{90}{C07 visit 2}}\end{picture} & 
\fbox{\includegraphics[width=2.8in, trim=0 16cm 0 3.5cm, clip]{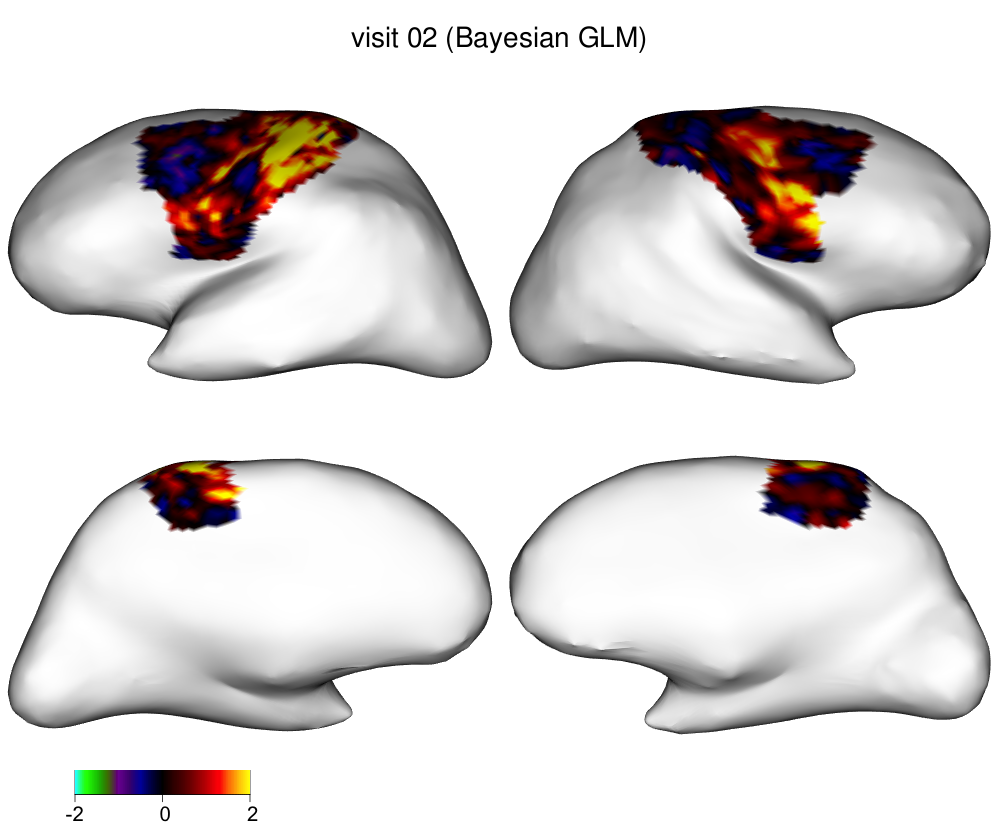}} &
\fbox{\includegraphics[width=2.8in, trim=0 13.5cm 0 3.5cm, clip]{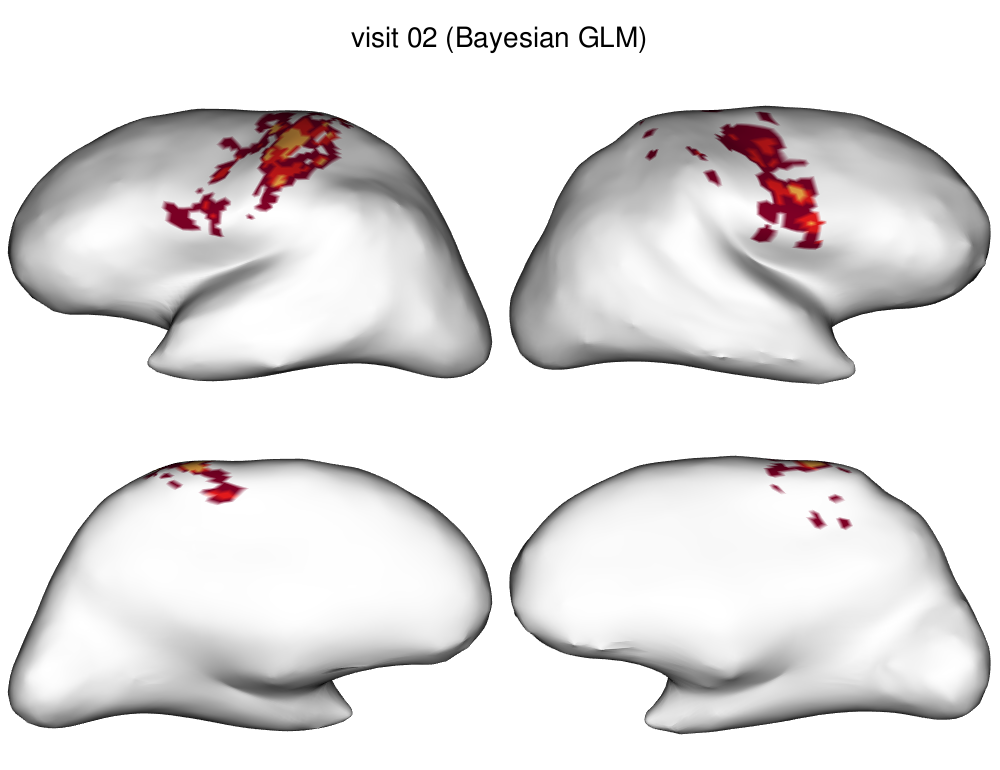}} \\[10pt]
\begin{picture}(0,60)\put(-5,30){\rotatebox[origin=c]{90}{C07 visit 3}}\end{picture} & 
\fbox{\includegraphics[width=2.8in, trim=0 16cm 0 3.5cm, clip]{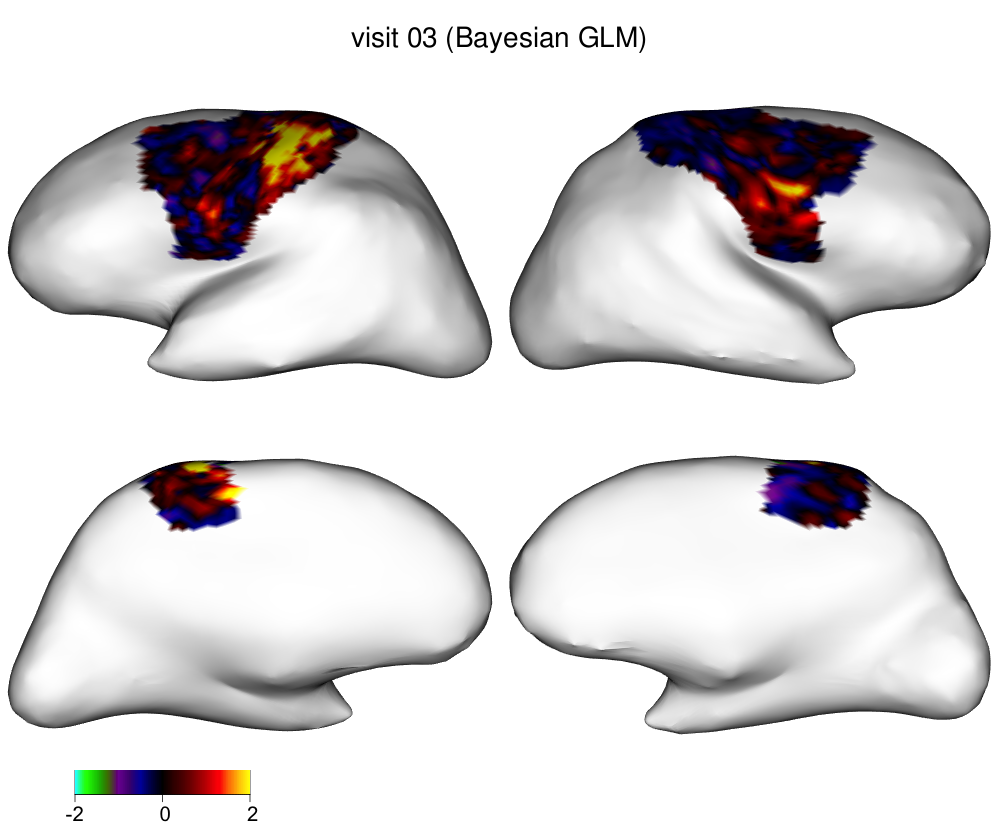}} &
\fbox{\includegraphics[width=2.8in, trim=0 13.5cm 0 3.5cm, clip]{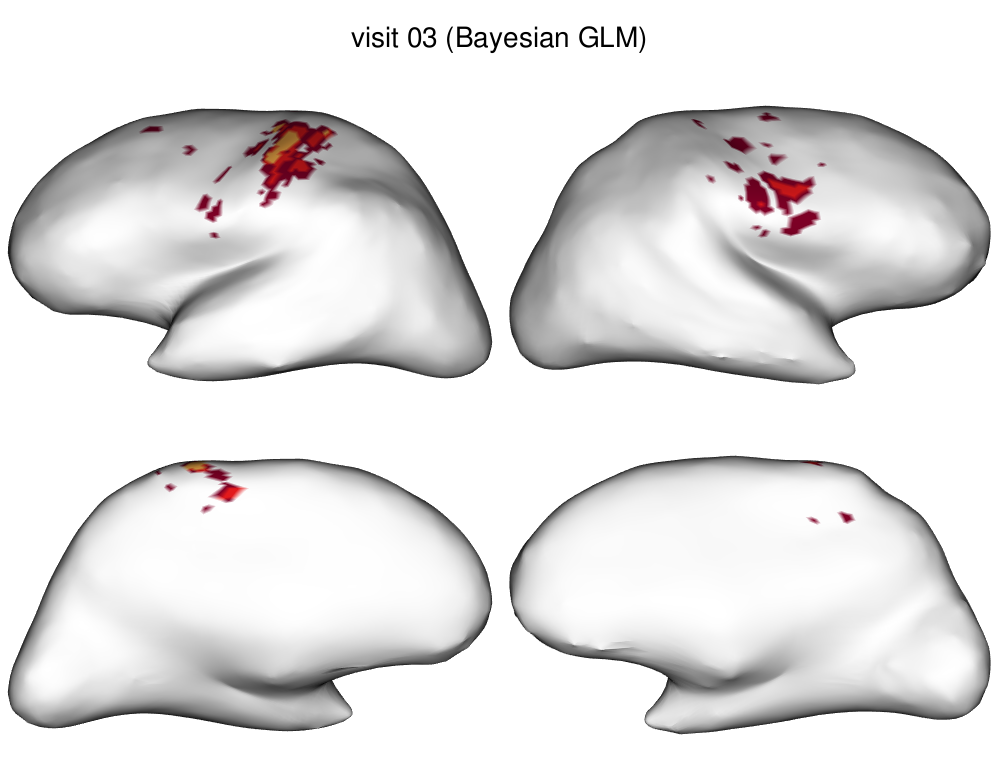}} \\[10pt]
& \multicolumn{2}{c}{\large \textbf{ALS Participant}} \\[5pt]
\begin{picture}(0,60)\put(-5,30){\rotatebox[origin=c]{90}{A26 visit 4}}\end{picture} & 
\fbox{\includegraphics[width=2.8in, trim=0 15cm 0 2.5cm, clip]{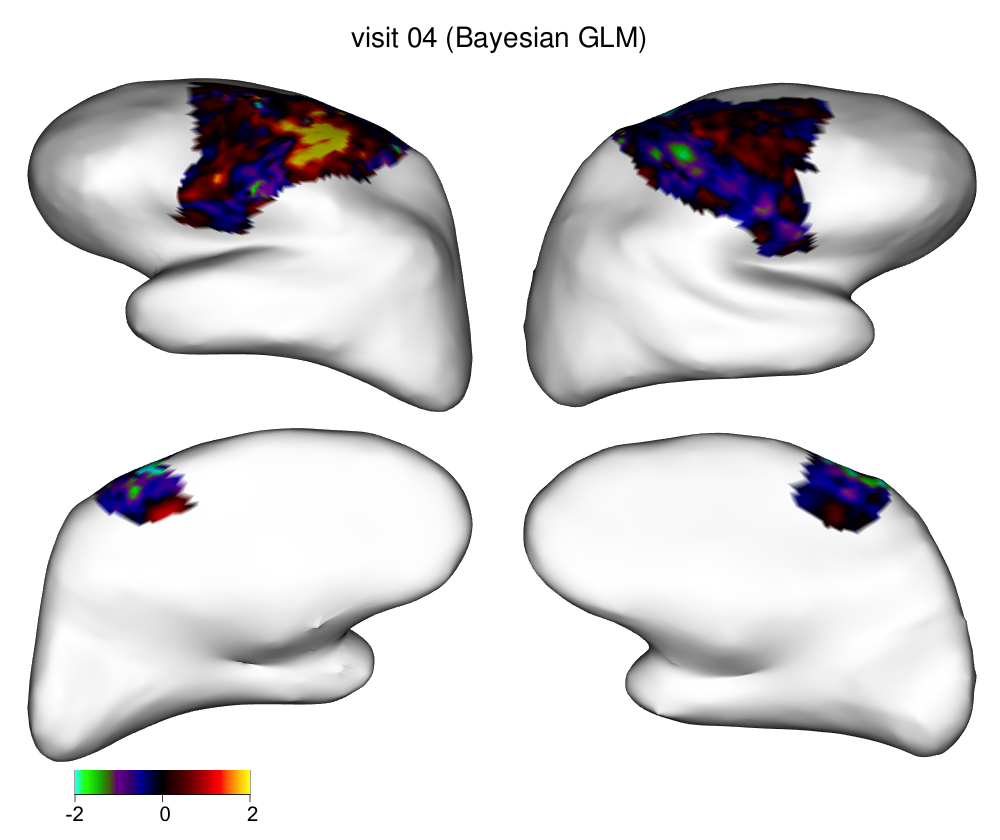}} &
\fbox{\includegraphics[width=2.8in, trim=0 12.5cm 0 2.5cm, clip]{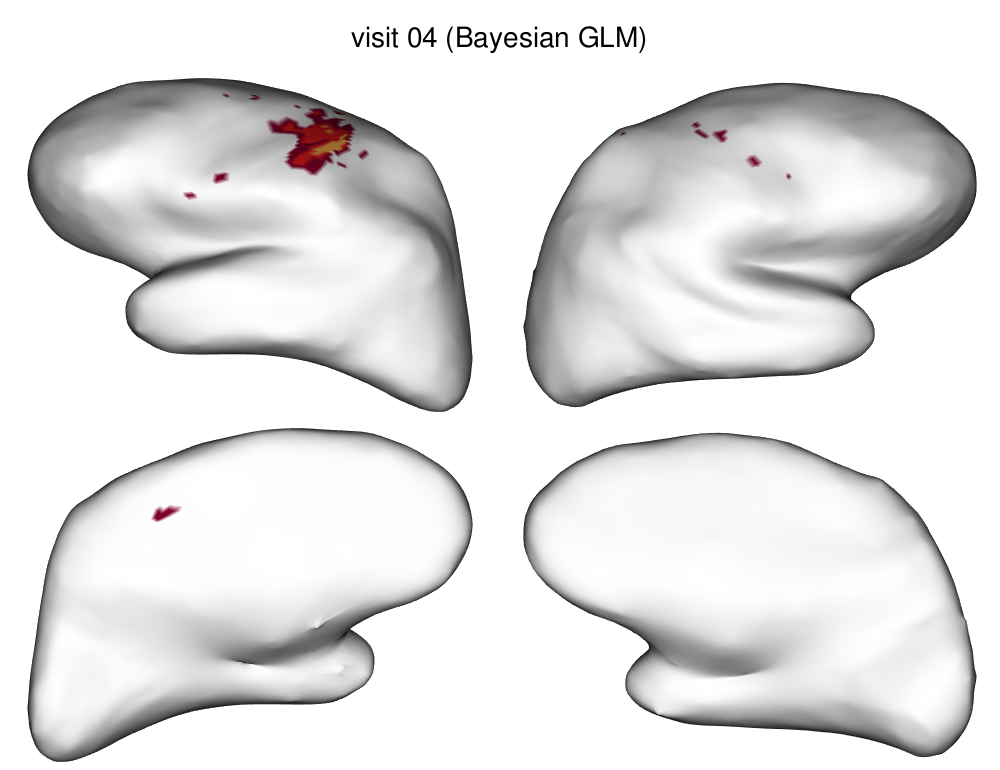}} \\[10pt]
\begin{picture}(0,60)\put(-5,30){\rotatebox[origin=c]{90}{A26 visit 5}}\end{picture} & 
\fbox{\includegraphics[width=2.8in, trim=0 15cm 0 2.5cm, clip]{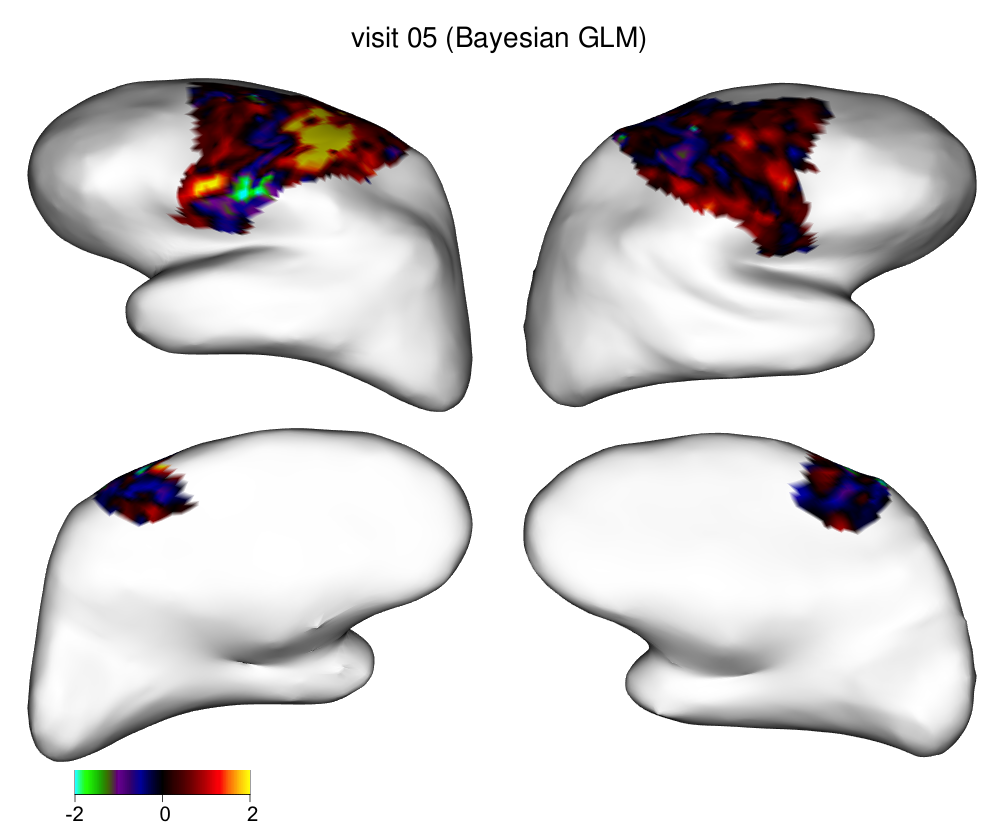}} &
\fbox{\includegraphics[width=2.8in, trim=0 12.5cm 0 2.5cm, clip]{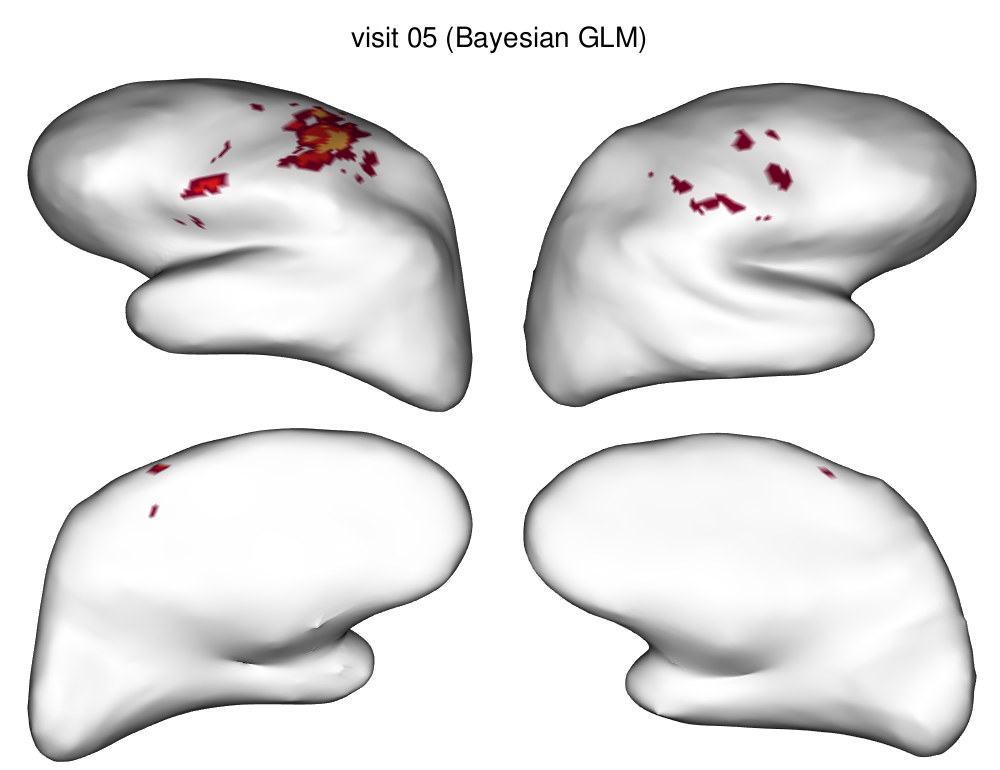}} \\[10pt]
\begin{picture}(0,60)\put(-5,30){\rotatebox[origin=c]{90}{A26 visit 6}}\end{picture} & 
\fbox{\includegraphics[width=2.8in, trim=0 15cm 0 2.5cm, clip]{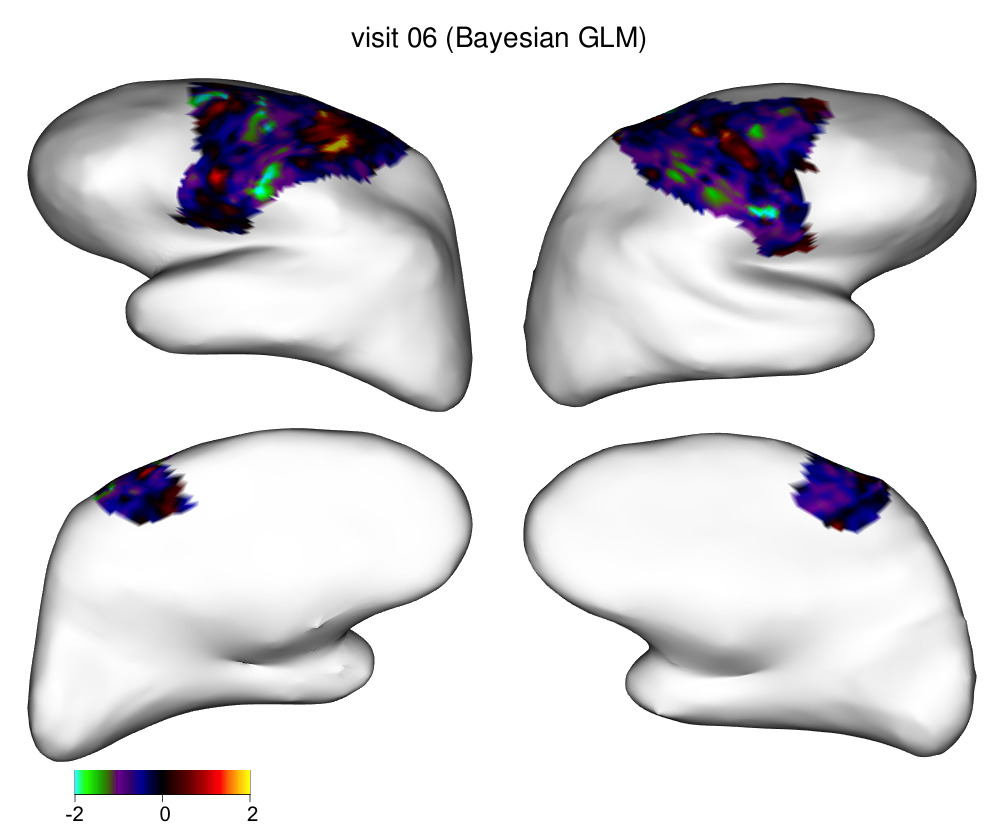}} &
\fbox{\includegraphics[width=2.8in, trim=0 12.5cm 0 2.5cm, clip]{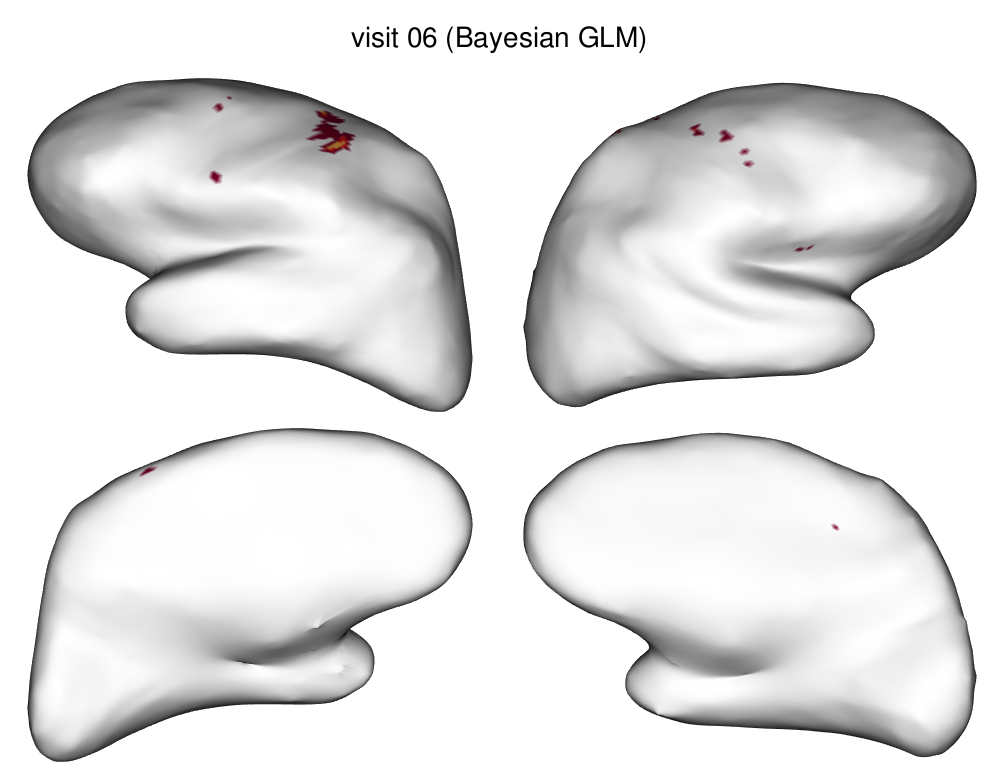}} \\[5pt]
& \includegraphics[width=1.3in]{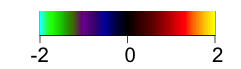} & \actlegendbay \\[5pt]
\end{tabular}
\caption{\small \textbf{Estimates and areas of activation during hand clenching in one HC and one ALS participant.} Areas of activation show binary maps of significance at three nested effect sizes, $\gamma=0\%$ (red plus orange plus yellow), $1\%$ (orange plus yellow), and $2\%$ (yellow). \textbf{Top panel:} The HC participant shows relatively stable patterns of activation over time. \textbf{Bottom panel:} The ALS participant shows noticeable changes between visits 4, 5 and 6 (their final three visits during the study). Between visits 4 and 5, the size of peak activation appears to increase somewhat, while between visits 5 and 6 the area peak activation virtually disappears. Similar patterns are observed in other ALS participants and appear to reflect dynamics of neurodegeneration in ALS.}
\label{fig:estimates}
\end{figure}
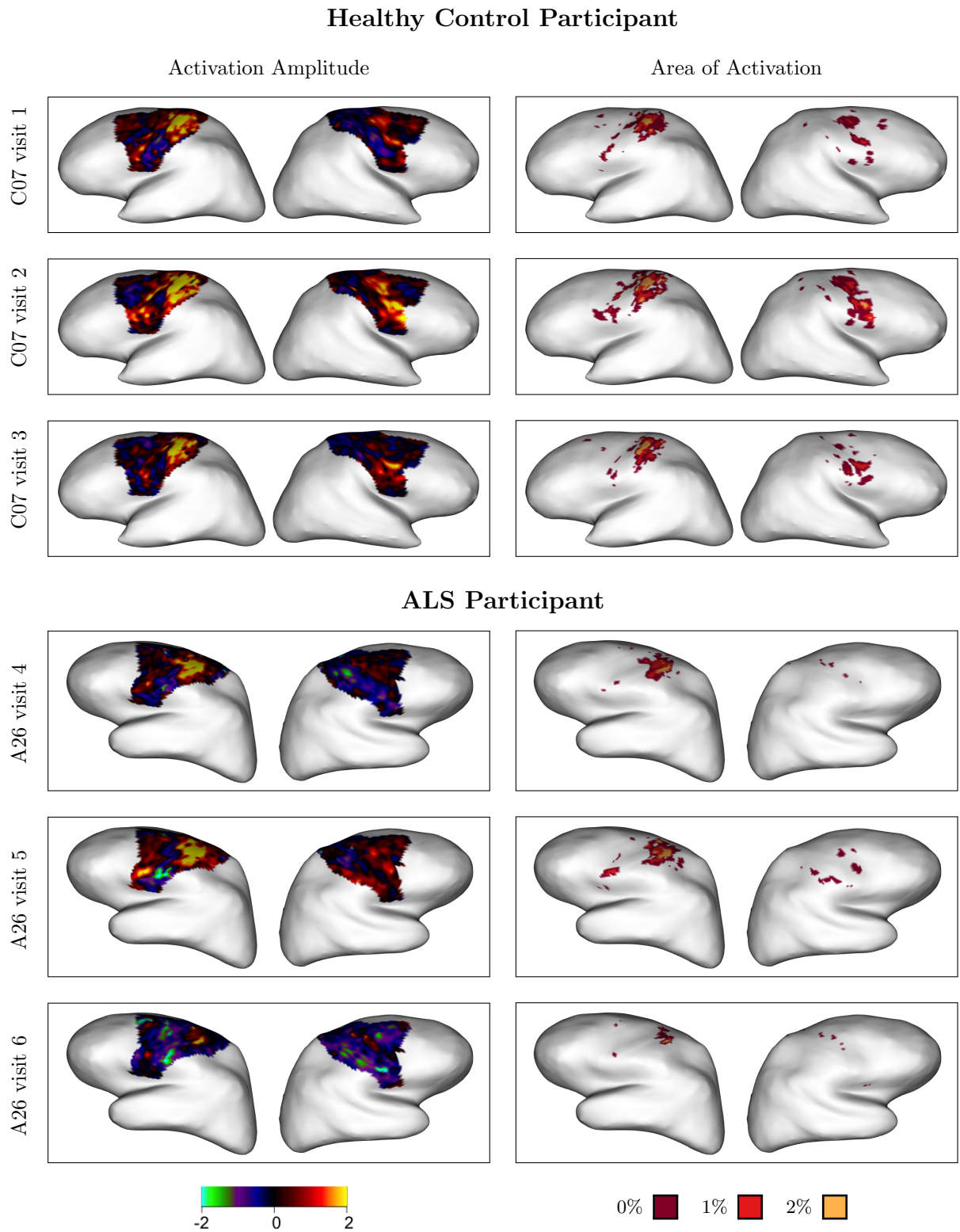

Fig.\ \ref{fig:estimates} displays longitudinal estimates and areas of activation for one HC and ALS participant. In the HC participant, the area of peak contralateral activation intensity was fairly consistent across visits. However, in the ALS participant, the peak intensified from visits 1 to 2 then shrunk markedly from visits 2 to 3. These patterns are reflected by the areas of activation, particularly $>1\%$ or $>2\%$ signal change. Similar patterns of increasing-then-decreasing activation over time were observed consistently in ALS participants. The participant shown, A26, was observed in Fig.\ \ref{fig:ALSFRS} to show a rapid functional decline over their final three visits (the ones displayed in Fig.\ \ref{fig:estimates}). We next examined this non-linear relationship between functional disability in ALS and size of motor activation.

\begin{figure}
    \centering
     \includegraphics[height=3in, page=3, trim=0 1.5cm 0 0, clip]{plots/lmer_activation_vs_diability_lh_excludeA04.pdf} 
     \includegraphics[height=3in, page=3, trim=6mm 1.5cm 0 0, clip]{plots/lmer_activation_vs_diability_rh_excludeA04.pdf}\\
    \includegraphics[width=2in]{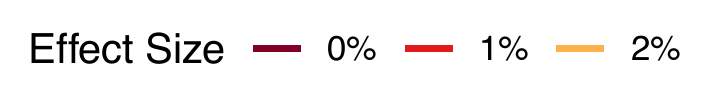}
        \caption{ \small \textbf{Relationship of total ALS disability to the size of contralateral and ipsilateral activation in response to right hand clenching.} Values are based on the mixed effects model given in Eqn. (\ref{eqn:lmer_ALS}). The colored dots on the left-hand side of each plot represent the mean size of activation for HC participants, based on the random intercept model given in Eqn. (\ref{eqn:lmer_HC}), with error bars showing one standard error around the mean.}
    \label{fig:mixed_effects_model_overall}
\end{figure}

Fig.\ \ref{fig:mixed_effects_model_overall} displays coefficient curves for the random intercept model given in Eqn. \ref{eqn:lmer_ALS_tot} relating \textit{Total Disability} in ALS to size of activation during right hand clenching. Each line represents the estimated relationship between disability and size of contralateral or ipsilateral activation at a given effect size. The colored dots on the left represent the corresponding mean activation in HCs, based on the model in Eqn. \ref{eqn:lmer_HC}.  The relationship is decreasing overall, but with marked non-linear features, most notably an increase in size of activation at moderate levels of disability. This suggests two distinct phenomena: a period of hyper-activation accompanied by an overall long-term decline in size of activation.  To better understand the drivers of these two distinct abnormal activation patterns, we then concurrently examined the relationship between size of activation and two separate measures of disability: \textit{Hand Motor Disability} and \textit{Other Disability}.
 
\begin{figure}
\begin{subfigure}[b]{1\textwidth}
    \centering
     \includegraphics[height=3in, page=1, trim=0 1.5cm 0 0, clip]{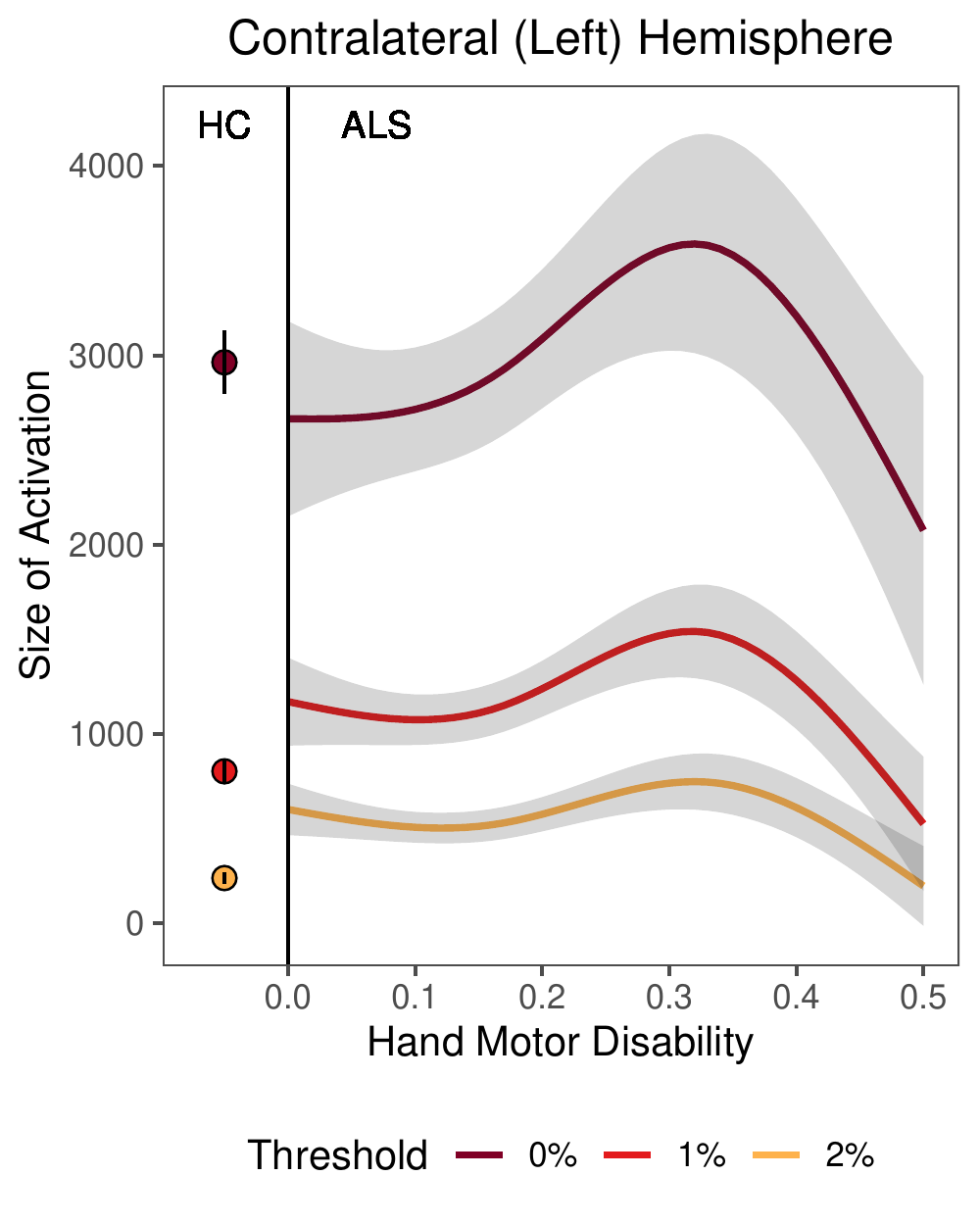} 
     \includegraphics[height=3in, page=1, trim=6mm 1.5cm 0 0, clip]{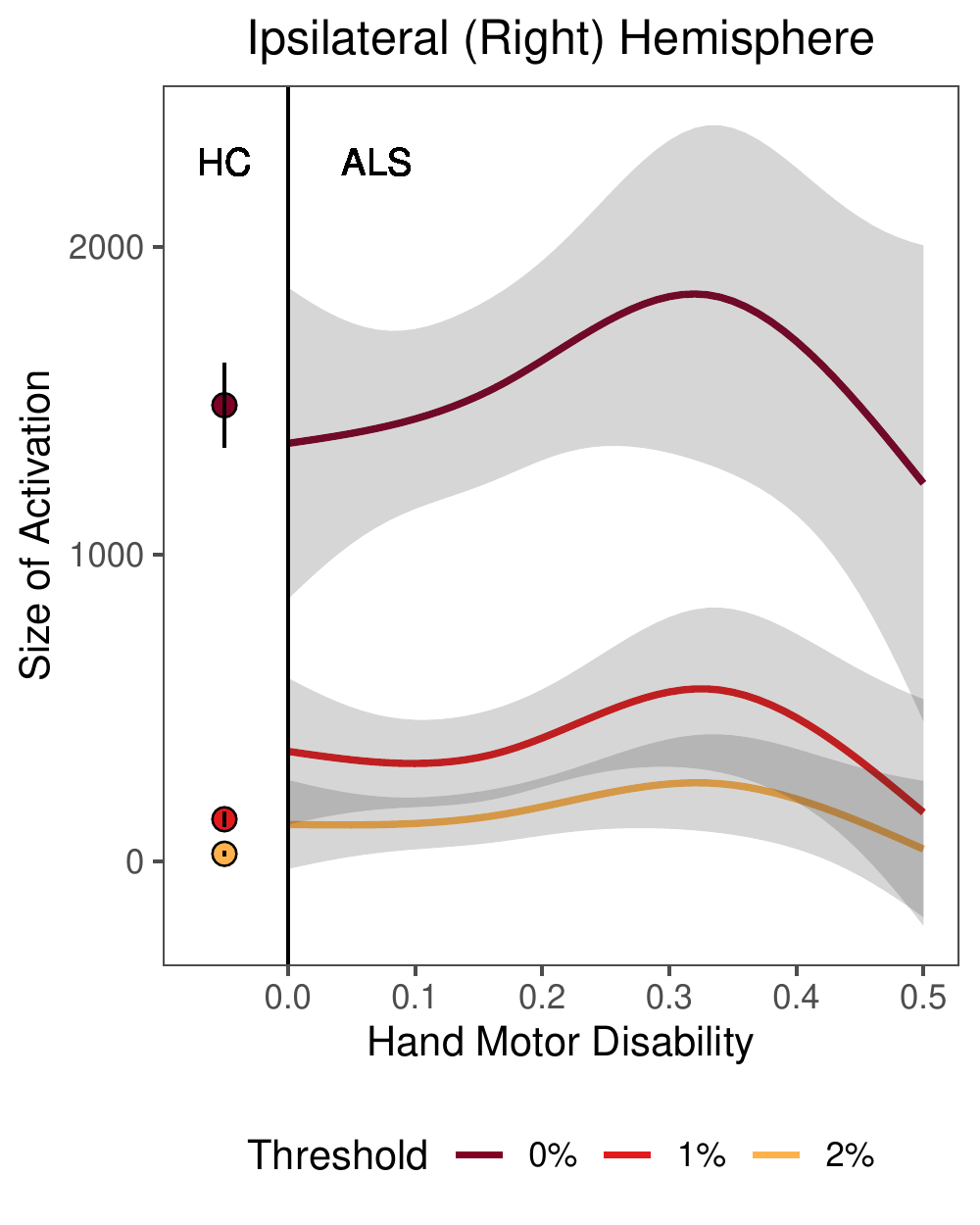} \\
    \includegraphics[width=2in]{plots/lmer_activation_vs_diability_legend.pdf}
    \caption{Relationship between Hand Motor Disability and size of activation during right hand clenching.\\[20pt]}
\end{subfigure}
\begin{subfigure}[b]{1\textwidth}
    \centering
    \includegraphics[height=3in, page=2, trim=0 1.5cm 0 0, clip]{plots/lmer_activation_vs_diability_lh_excludeA04.pdf} 
    \includegraphics[height=3in, page=2, trim=6mm 1.5cm 0 0, clip]{plots/lmer_activation_vs_diability_rh_excludeA04.pdf} \\
    \includegraphics[width=2in]{plots/lmer_activation_vs_diability_legend.pdf}
    \caption{Relationship between Other Disability and size of activation during right hand clenching.}
\end{subfigure}
\caption{ \small \textbf{Relationship between the size of activation and hand motor disability and other disability in ALS.} Values are based on the mixed effects model given in Eqn. (\ref{eqn:lmer_ALS}). The predictor variables are \textit{Hand Motor Disability} and \textit{Other Disability}. The range of the x-axes represent up to the 90th quantile of each predictor. Shaded bands show one standard error around the mean.  The colored dots on the left-hand side of each plot represent the mean size of activation for HC participants, based on the random intercept model given in Eqn.  (\ref{eqn:lmer_HC}), with error bars showing one standard error around the mean. The curves shown in panel (a) reveal an inverted-U-shaped relationship between size of activation and \textit{Hand Motor Disability}; panel (b) shows decreasing size of activation associated with other aspects of disability.}
\label{fig:mixed_effects_model}
\end{figure}

Fig.\ \ref{fig:mixed_effects_model} displays coefficient curves estimating the relationship between size of activation and these two separate disability measures, based on the random intercept model given in Eqn. \ref{eqn:lmer_ALS}.  Each line represents the estimated relationship between the predictor shown on the x-axis and size of activation at a given effect size, holding the other predictor constant at zero.  The relationship between size of activation and \textit{Hand Motor Disability} exhibits a clear inverted U-shaped relationship: at moderate levels of disability, there is a sharp increase over the normal size of activation in HCs, but with further disability there is a rapid decline to abnormally low levels. Considering only measures of disability not associated with hand function (\textit{Other Disability}), there is a purely declining relationship between disability and size of activation.  This suggests that hyper-activation during hand clenching is specifically associated with declining hand motor function. 

Fig.\ \ref{fig:mixed_effects_model} shows two additional effects. First, the size of activation when \textit{Hand Motor Disability} or \textit{Other Disability} reach higher levels is nearly zero at effect sizes of $1\%$ and $2\%$ in both hemispheres, suggesting near complete loss of neuronal activation as the processes of neurodegeneration associated with ALS disease progression continues. Second, ALS participants with very low levels of disability (\textit{Hand Motor Disability} and \textit{Other Disability} both equalling 0) exhibited slightly elevated contralateral and ipsilateral activation at effect sizes of $1\%$ and $2\%$, relative to HCs. This suggests that patterns of hyper-activation may occur very early in the ALS disease process, even prior to the onset of measurable disability.


\begin{figure}
\begin{subfigure}[b]{0.38\textwidth}
    \centering
    \includegraphics[width=1\textwidth, page=2]{plots/ALSFRS_fastslow.pdf}
    \caption{ALS Progression Rates}
\end{subfigure}
\begin{subfigure}[b]{0.61\textwidth}
    \includegraphics[width=1\textwidth]{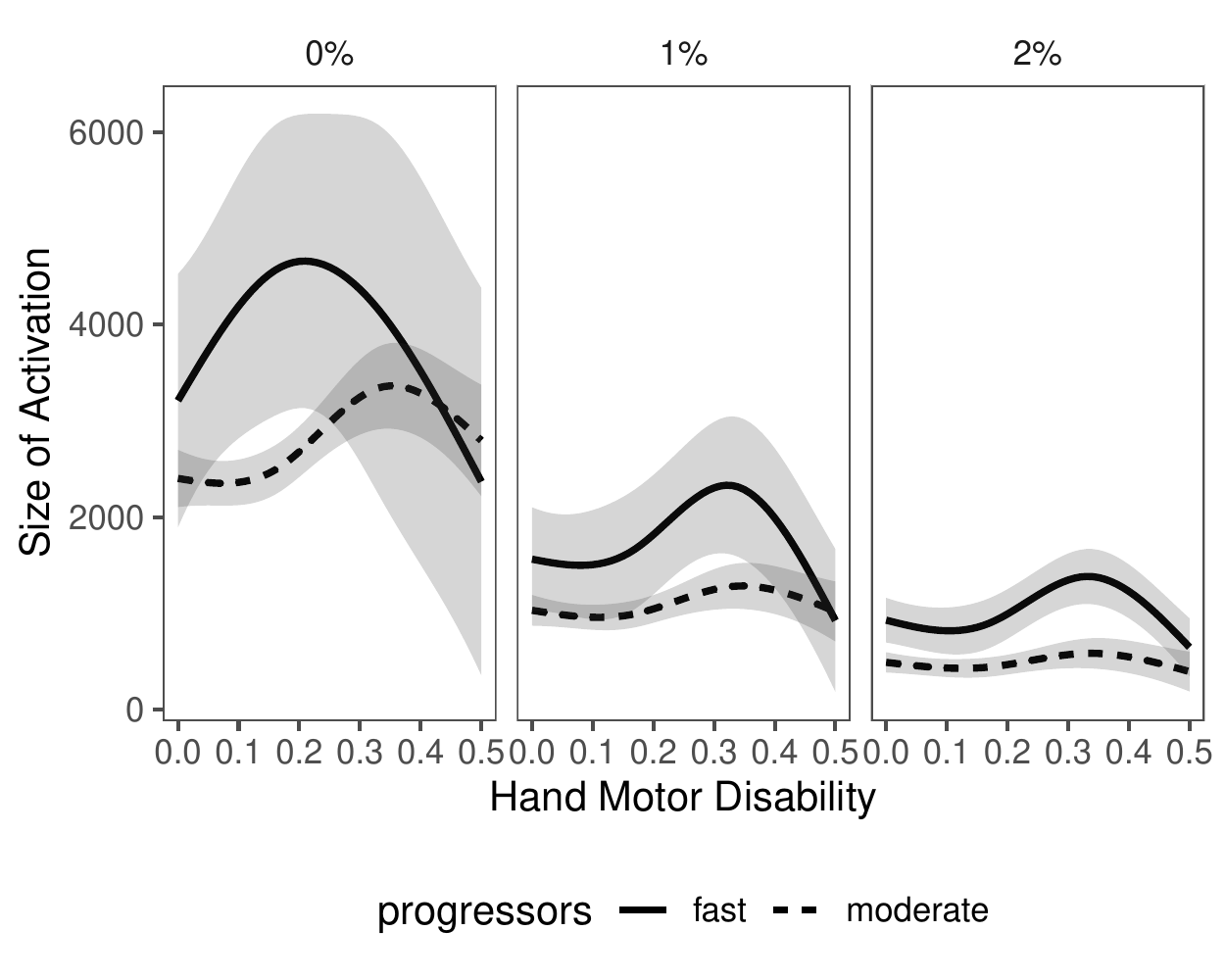}\\[-4pt]
    \caption{Activation Trajectories by ALS Progression Rate}
\end{subfigure}
    \caption{ 
        \small \textbf{Relationship between activation size and hand motor disability by ALS progression rate.} \textbf{(a)} Disability progression rate of ALS subjects and classification. \textbf{(b)} Coefficient curves for the size of contralateral activation in response to right hand clenching, based on the model in Eqn. \ref{eqn:lmer_ALS} stratified by progression rate. Shaded bands show one standard error around the mean. Different panels correspond to the effect sizes $0\%$, $1\%$ and $2\%$ signal change. The overall shape is consistent with Fig.\ \ref{fig:mixed_effects_model}. However, fast progressors tend to peak higher and have higher baseline size of activation than moderate progressors. Furthermore, for the $0\%$ effect size, the fast progressors tend to peak at an earlier stage of hand motor disability. Corresponding plots for ipsilateral activation are given in the Supplementary Materials and show similar patterns.}
    \label{fig:mixed_effects_model_fastslow}
\end{figure}

Fig.\ \ref{fig:mixed_effects_model_fastslow} displays results stratified by progression rate. The overall relationships with \textit{Hand Motor Disability} are consistent with the inverted U-shaped trajectories observed in Fig.\ \ref{fig:mixed_effects_model}. However, compared with moderate progressors, fast progressors exhibit three noteworthy differences: first, they tend to have greater baseline size of activation (when \textit{Hand Motor Disability} is near zero); second, they tend to peak higher; third, for $0\%$ effect size, they tend to peak at an milder level of \textit{Hand Motor Disability}. Corresponding plots for ipsilateral activation show similar patterns (see Supplementary Sec. \ref{app:more_result_figures}). Supplementary Fig.\ \ref{fig:mixed_effects_model_fastslow_other} displays the relationship between size of activation and \textit{Other Disability} for fast and moderate progressors. Fast progressors also exhibit greater enlargement of activation at low levels of \textit{Other Disability} and decline faster as a function of disability. Since fast progressors experience a given level of disability earlier following symptom onset, these findings suggest that the neuronal manifestations of fast-progressing disability may outpace or even precede clinical disability.

\section{Discussion}
\label{sec:discussion}

Using a rich longitudinal task fMRI dataset and an advanced cortical surface spatial Bayesian modeling approach, in this study we observed complex longitudinal trajectories of neurodegeneration related to disability in ALS. Below, we discuss these findings in the context of previous literature and suggest directions for future research into neurodegeneration in ALS. We also discuss the importance of analysis and processing choices for longitudinal modeling and other settings where accurate individual-level brain measures are needed, including biomarker discovery. Finally, we note several limitations of this study.


\subsection{Complex trajectories of motor activation}

We observed three consistent patterns of change in motor activations. First, we observed increased motor activation elicited by a simple hand clench task associated with mild hand motor disability. Second, as patients experience further disability, we observed a sharp reduction in the size of activation. Third, we also observed reduced activation associated with other aspects of disability. These latter two observations suggest major neuronal degeneration at more advanced levels of disability. 

These effects are most clearly observed contralaterally, but ipsilateral activations exhibit similar patterns through the disease process. Note that some level of ipsilateral activation is expected in lateral motor tasks \citep{Konrad:2002hl}, as we observe in HCs and in ALS subjects with very low levels of disability.  

We also observed that, compared with moderate progressors, fast progressors had increased hyper-activation occurring at an earlier stage of hand motor disability. These findings indicate that certain patterns of neurodegeneration may precede or outpace the rate of clinical disability. This builds on the finding of \cite{Douaud:2011hb}, who observed differential functional connectivity patterns in ALS, in which individuals with slower progression rates more closely resembled HCs.  

These findings substantially advance prior understanding of motor neurodegeneration in ALS. Such discoveries were made possible by our rich longitudinal study that followed ALS participants over the course of disease progression, from mild to more advanced disability. Following participants until they were no longer able to undergo scanning enabled us to discover a late decline in activation following the initial increase associated with milder disability. While increased motor activation in ALS has been noted in previous studies \citep{Schoenfeld:2005eh, Poujois:2013hn, Konrad:2002hl}, to our knowledge this subsequent decline has not been observed in prior literature. This, along with the decline we observed associated with non-hand disability, likely reflects the drastic motor neuron loss known to occur in ALS \citep{Subramaniam:2019uq}.


\subsection{Does increased activation reflect compensation or loss of inhibition?}

Our observation of increased motor activation associated with mild hand disability is consistent with several studies over the past two decades using fMRI to reveal motor cortex changes in ALS. However, we observe more complex longitudinal trajectories, which suggest that the initial hyper-activation may not be fully attributable to compensation, as suggested by earlier work. Previously, \cite{Konrad:2002hl}, \cite{Schoenfeld:2005eh} and \cite{Poujois:2013hn} observed enlarged activations during motor tasks. \cite{Poujois:2013hn} related this to disease progression rate but with cross-sectional data. Such hyper-activation has been typically interpreted as a \textit{compensatory} process enabling the patient to complete the task in ALS \citep{Konrad:2002hl, Schoenfeld:2005eh}, building from findings in healthy aging \citep{Park:2003jw,Fitzhugh:2019jy}. In ALS, however, increased activation could alternatively be attributed to \textit{diminished inhibitory signaling} within the brain \citep{Poujois:2013hn}, which is known to occur in ALS.

Our findings point to loss of inhibition as a likely contributor to the hyper-activation associated with mild disability. First, we observe similar patterns of early hyper-activation \textit{ipsilaterally}, which is more suggestive of loss of inhibition. Second and perhaps more notably, the differential trajectories observed across fast and moderate progressors conflict with a pure compensation model, since these two groups are experiencing hyper-activation at different levels of motor ability. Instead, through the lens of loss of inhibition, the earlier and more extreme hyper-activation observed in fast progressors may simply reflect more extreme and rapid neurodegeneration of inhibitory pathways at this stage of the disease process in fast progressors. It is possible that such an initial loss of inhibition, followed by later hypo-activation with additional disease burden, are part of a single process of motor neuron loss in ALS \citep{Subramaniam:2019uq}.

Prior studies employing other neuroimaging modalities can also shed light on this question. Transcranial magnetic stimulation (TMS) has been used to probe inter-hemispheric communication, specifically early intracortical inhibition (ICI). \citep{Zanette:2002un} observed that abnormal ICI developed early in the ALS disease process and continued to further degrade with disease progression. Magnetic resonance spectroscopy (MRS) has revealed decreased endogenous gamma-Aminobutyric acid (GABA) \citep{Lloyd:2000kj, Foerster:2013il}, a principal inhibitory neurotransmitter, in individuals with ALS. Although another study of GABA by \cite{Blicher:2019iv} did not see such a change, it was based on a small sample and voxel size, which may have led to insufficient statistical power. Studies of functional connectivity have produced somewhat conflicting findings, reporting both increased and decreased connectivity \citep{Mohammadi:2009es,JelsoneSwain:2010gj,Verstraete:2010gr}. \cite{Douaud:2011hb} suggested a possible explanation for this. Using functional and structural connectivity, they observed initial loss of inter-hemispheric inhibition giving rise to increased trans-hemispheric connectivity. This eventually wanes due to loss of the neurons responsible for the random fluctuations producing functional connectivity, with connectivity greatly diminishing and eventually falling below baseline as the patient progresses. This is consistent with our findings of an inverted U-shaped activation trajectory. In sum, there is strong prior evidence for loss of inhibition in ALS from studies employing TMS, MRS, and connectivity.

\subsection{Producing accurate individual-level measures of brain function}

Our discoveries were facilitated by a sophisticated analysis approach designed to provide accurate and reliable individual-level measures of task activation, in contrast with conventional task fMRI analysis methods that tend to exhibit poor reliability in individuals \citep{elliott2020test}. This is important for longitudinal studies, as well as other settings where robust and reliable individual-level measures are needed. Two key aspects of our approach were: 1) performing analysis in subject-specific surface space and 2) adopting a novel longitudinal surface-based spatial Bayesian GLM.

\subsubsection{Performing analysis in subject surface space}

We constructed subject-specific surface templates, enabling us to preserve anatomical features and avoid potential normalization issues in individuals with neurodegenerative disease. \cite{Eloyan:2014hz} gave a striking example of such issues in multiple sclerosis: they found that standard normalization methods resulted in a large proportion of white matter lesions being relocated outside of white matter. Even if subjects are aligned anatomically, they may not align functionally due to individual differences, reducing the utility of voxel-level comparisons in standard space \citep{dubois2016building}. Issues of misalignment and distortion can be mitigated by performing analyses in subject space. Performing surface-based analysis also helped enhance specificity by avoiding blurring across tissue classes or distinct areas of the cortex \citep{brodoehl2020surface}. 

Subject-space analyses require aggregating and comparing subject-level results in a different way than traditional voxel-wise comparisons. In our analysis, we used size of activation as the basis for examining disease trajectories and group differences, facilitated by the robust areas of activation produced by the spatial Bayesian GLM. Several alternatives are possible: \cite{Gupta:2010ff} used spatial properties of activation patterns for predictive modeling, while \cite{Stern:2009ex} used topographical analysis. Both approaches outperformed voxel-wise group-difference analysis in terms of discovering disease effects. Notably, \cite{Stern:2009ex} found HCs to exhibit high spatial heterogeneity in activation, suggesting that performing analysis in subject space may be beneficial more generally. 

\subsubsection{The advantages of longitudinal spatial Bayesian modeling}

Surface-based spatial Bayesian modeling of task activation has been previously validated and shown to produce more accurate and robust activations by leveraging spatial dependencies \citep{mejia2020bayesian}. We proposed a novel longitudinal extension, which we validated by examining the stability of results for HC participants over time (see Supplementary Section \ref{app:validation}). This extension has important advantages over the single-session model. By pooling information across sessions to estimate model parameters, it produces more accurate estimates and areas of activation and reduces longitudinal noise, enhancing subsequent longitudinal analysis. This modeling framework can be extended to non-longitudinal contexts, such as cross-sectional studies with multiple sessions per subject or small group studies.

\subsection{The next frontier: Biomarker discovery}

Motor cortex changes hold potential as a possible biomarker for ALS disease diagnosis and progression. Brain signatures may signal early loss of inhibition, even in pre-symptomatic disease, that might serve as a biomarker for ALS diagnosis. Changes (structural and functional connectivity) due to the ALS disease process \citep{Lee:2017fh} have been observed in pre-symptomatic carriers of C9orf72 (a genetic defect linked to ALS \citep{Renton:2011di}. Our findings suggest the possibility of very early hyper-activation above a scientifically meaningful effect size. Future work should assess hyper-activation and loss of inhibition as potential biomarkers for pre-clinical ALS. Declining motor activation following a period of over-activation may also serve as a marker of a change to a later phase of disease progression. 

Clinical trials employing clinical outcome measures are typically lengthy and expensive. Development of a brain biomarker of ALS reflecting enhanced understanding of pathophysiology would have major implications for clinical trials and therapies \citep{Gordon:2010gq} by making them more efficient and effective \citep{Turner:2013dp}. For example, a trial of a drug targeting abnormal cortical excitability could employ such a biomarker to select only subjects that are in early phases of motor neurodegeneration. Perhaps more importantly, such a biomarker could also be used to monitor and guide treatment, by escalating drug dosage until suppression of hyper-excitability is achieved or decline in activation over time is arrested.  Vitally, a biomarker that is sensitive to cortical changes could produce evidence of drug intervention efficacy prior to manifestation in measurable clinical decline \citep{Turner:2009fp}.

\subsection{Study limitations}

As with other rare neurological diseases, neuroimaging studies of ALS are difficult to execute. Given the rarity and typically short survival time characterizing ALS, sample sizes tend to be smaller, especially for single-site studies. While consortia studies are taking place, these are currently focused on structural and resting state investigation \citep{Bharti:2020exa}. While the sample size in our study is a limitation, a literature search suggests that our study is one of the largest, if not the largest \citep{Trojsi:2020el,Castelnovo:2020fi}, longitudinal studies investigating BOLD activation in ALS. We sought to mitigate the negative impacts of small sample size by employing a sophisticated analysis approach, thus avoiding the power issues associated with a classical massive univariate analysis. 

One limitation common to neuroimaging studies of ALS is the requirement that ALS participants must be capable of lying in a prone position for some length of time during MRI scanning. While this constraint can be lessened by allowing the individual to take breaks and sit up as needed, it still imposes a bias on neuroimaging studies of ALS \citep{vanderBurgh:2020er}, since they are limited to only those that can tolerate the MRI. A seated-position MRI would allow patients with more disability to participate, but these systems have greatly reduced magnetic field strengths of 0.25T (G-scan; Esaote SpA, Genoa, Italy) or 0.60T (FONAR Melville, New York, USA). To mitigate this source of bias, we examine cortical activation as a function of physical disability, as opposed to time since symptom onset, which helps account for the bias toward subjects with lower disease burden. Even so, our study is limited in terms of the range of disease severity we were able to observe.

In our analysis, we only considered activation within the motor mask (Supplementary Fig.\ \ref{fig:mesh_ALS}), so activation in other areas cannot be observed. It is conceivable that the enlarged areas of activation we observe in ALS may expand beyond the mask used in this study, resulting in possible underestimation of size of activation for some subjects. Additionally, our analysis did not include subcortical or cerebellar areas. Future work should focus on analyzing longitudinal trajectories of motor activation in ALS across the entire cortex and within relevant subcortical and cerebellar regions. Finally, our analysis did not consider atrophy over the course of ALS disease progression. Atrophy may help to partly explain the dramatic drop in motor activation occurring with high disability but would not explain increased activation at lower levels of disability. Future research should aim to development models to incorporate atrophy.

\section{Conclusion}

In this paper, we adopted a sophisticated longitudinal surface-based Bayesian analysis approach to analyze a rich longitudinal fMRI study of ALS.  In this study, individuals with ALS and matched healthy controls were observed regularly for 1-2 years or longer. Our analyses revealed a complex trajectory of cortical activation during a simple motor hand clench task: activation initially spreads within contralateral and ipsilateral motor areas, but with additional disease burden activations sharply diminish and eventually nearly disappear. We observed systematic differences based on clinical progression rate, with fast progressors exhibiting more extreme effects earlier in the disease process. The nuances of these findings suggest that initial hyper-activation---observed in earlier studies but assumed to be due to functional compensation---is likely due to a loss of inhibitory signals forming part of a larger process of neuronal decay and death. These discoveries were made possible by pairing a rich longitudinal fMRI dataset with a surface-based spatial Bayesian modeling approach capable of identifying activations in individuals over time with high accuracy and power. Our study establishes that this advanced statistical approach furthers the study of neurodegenerative disease and is promising for the study of other time-varying processes such as development and aging. The surface-based spatial Bayesian GLM is implemented in a user-friendly R package, \texttt{BayesfMRI}.

\section*{Funding}

This work was supported by the National Institute of Biomedical Imaging and Bioengineering at the National Institutes of Health (R01EB027119 to A.F.M.), the National Institute of Neurological Disorders and Stroke at the National Institutes of Health (R01NS052514, R01NS082304 to R.C.W.), and the Department of Radiology at the University of Michigan (BRS Award to R.C.W).

\section*{Acknowledgements}

This study would not have been possible without the generous commitment of our participants with amyotrophic lateral sclerosis and their families. These patients and their families committed years to this study during an exceedingly difficult time in their lives, especially true for a longitudinal study such as ours. This work is dedicated to these patients.

\bibliography{mybib.bib}
\bibliographystyle{apalike}

\appendix

\renewcommand\thefigure{\thesection\arabic{figure}}    
\renewcommand\thetable{\thesection\arabic{table}}    
\setcounter{figure}{0} 
\setcounter{table}{0} 
\setcounter{equation}{0}
\newpage
\section{Participant Details}\label{app:participants}
\setcounter{page}{1}

All participants with amyotrophic lateral sclerosis presented with limb onset only. We did not genotype any participants. No participants had frontotemporal dementia. We did not screen, though for any type of cognitive impairment. Healthy control participants in general were matched on sex and age as a group, with no statistically significant differences in age.

\begin{table}[h]
\begin{center}
\begin{tabular}{c c c c c c c }
\hline 
Participant & Age  &  Age  & Sex &  ALSFRS-R  & ALSFRS-R  & Number of  \\
 & first visit (yr) &  last visit &  &  first visit &  last visit & Visits \\
\hline 
A04 & 58.7 & 63.1 & M & 44 & 43 & 10\\
A06 & 53.4 & 54.2 & M & 43 & 38 & 3\\
A08 & 64.9 & 65.1 & M & 29 & 29 & 3\\
A11 & 57.8 & 58.4 & M & 42 & 39 & 4\\
A14 & 55.1 & 57.2 & M & 42 & 35 & 7\\
A18 & 61.3 & 63.0 & F & 42 & 35 & 7\\
A19 & 67.8 & 69.2 & F & 43 & 31 & 7\\
A21 & 66.8 & 67.2 & M & 40 & 30 & 4\\
A23 & 56.7 & 57.0 & M & 39 & 31 & 3\\
A25 & 54.6 & 54.8 & M & 40 & 37 & 3\\
A26 & 58.3 & 59.6 & M & 46 & 27 & 6\\
A30 & 63.2 & 64.1 & M & 35 & 24 & 4\\
A31 & 65.8 & 66.6 & F & 26 & 25 & 4\\
A32 & 47.1 & 48.2 & M & 36 & 30 & 5\\
A33 & 55.3 & 56.2 & F & 46 & 44 & 5\\
A34 & 56.0 & 56.2 & M & 38 & 31 & 3\\
\hline \\
\end{tabular}
\end{center}
\caption{\small ALS Participants}
\end{table}

\begin{figure}
    \centering
    \includegraphics[width=6in]{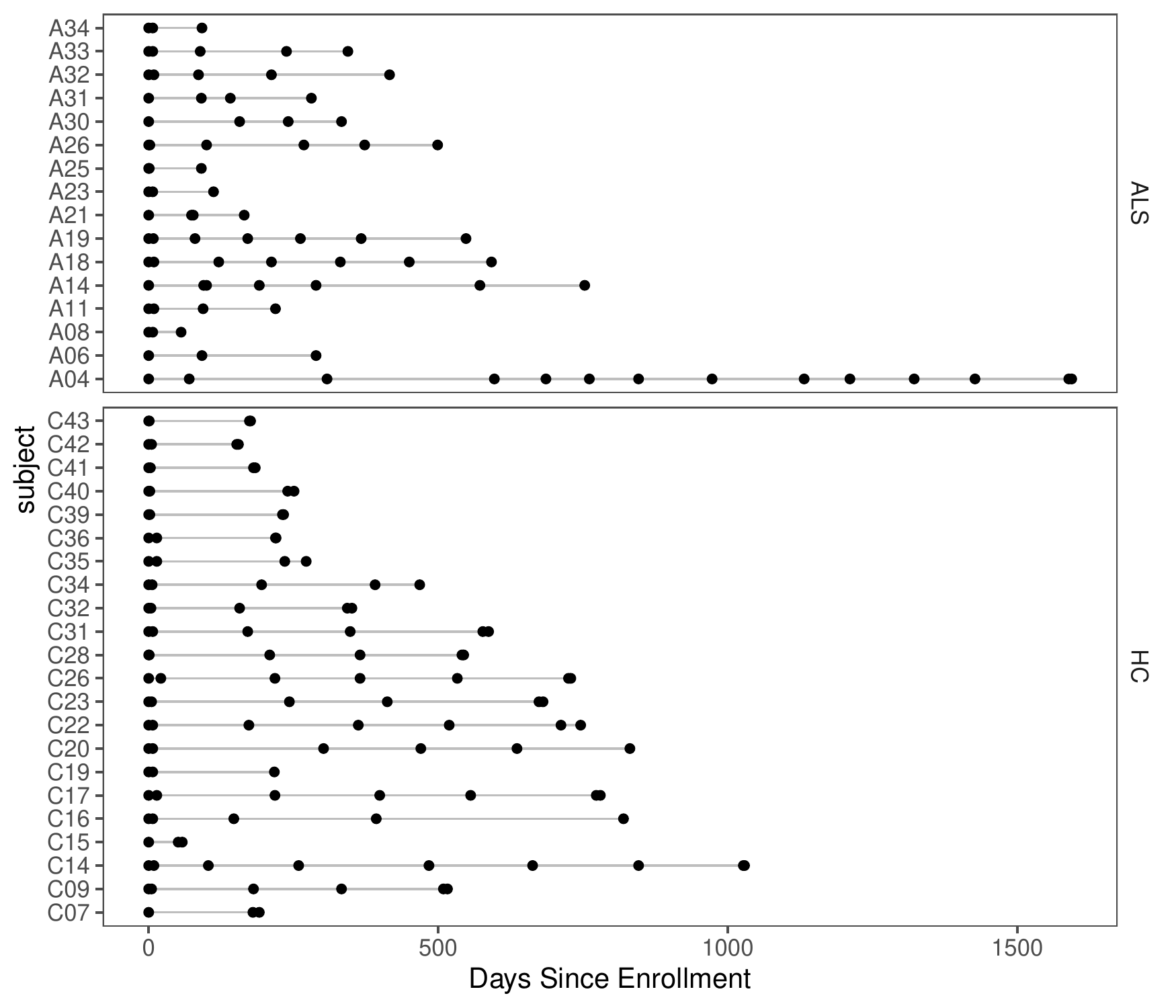}
    \caption{\small \textbf{Visit timing for each participant in the study.} Each dot represents a MRI session visit. Note that for many participants, the first two visits occurred in quick succession and appear overlapping on the plot.}
    \label{fig:visit_timing}
\end{figure}

\newpage
\section{Processing and Analysis Details}\label{app:processing}



BOLD time-series were projected to the unique cortical surface of each individual participant. For each research visit, data were processed with a blend of the SPM (Statistical Parametric Mapping, Verson 12, Release 7219, University College London) \citep{spm_book} software package, FSL (Functional Magnetic Resonance Imaging of the Brain Software Library, Version 6, Oxford University) \citep{Jenkinson:2012dj}, ANTs (2.3.1, University of Pennsylvania) \citep{Avants:2011kk}, and FreeSurfer (Version 6.0, Harvard and Mass General Hospital) \citep{Fischl:2012el}, and finally the Human Connectome Project Workbench (Version 1.2.3). Time-series data were slice-time corrected and realigned (2 passes). The lower resolution $T_1$-weighted image was co-registered to the mean realigned BOLD image. The high-resolution $T_1$-weighted image was then co-registered to the resulting co-registered low-resolution image. Next, non-uniformity correction was applied to the original $T_1$-weighted images using ANTs' N4 algorithm (variant of the N3 algorithm, nonparameteric nonuniform normalization). For each individual, the resulting bias field corrected $T_1$-weighted images of all sessions were then fed into ANTs' \texttt{antsMultiVariateTemplateConstruction2.sh} pipeline to create a participant template that is spatially unbiased to the orientation of the input images. Only rigid-body registrations were used for the template building. The participant's template image was then processed through FreeSurfer, including edits for brain mask, to result in a model of the pial surface and a corresponding spherical surface for the participant. 

\begin{figure}[H]
    \centering
    \includegraphics[width=5in]{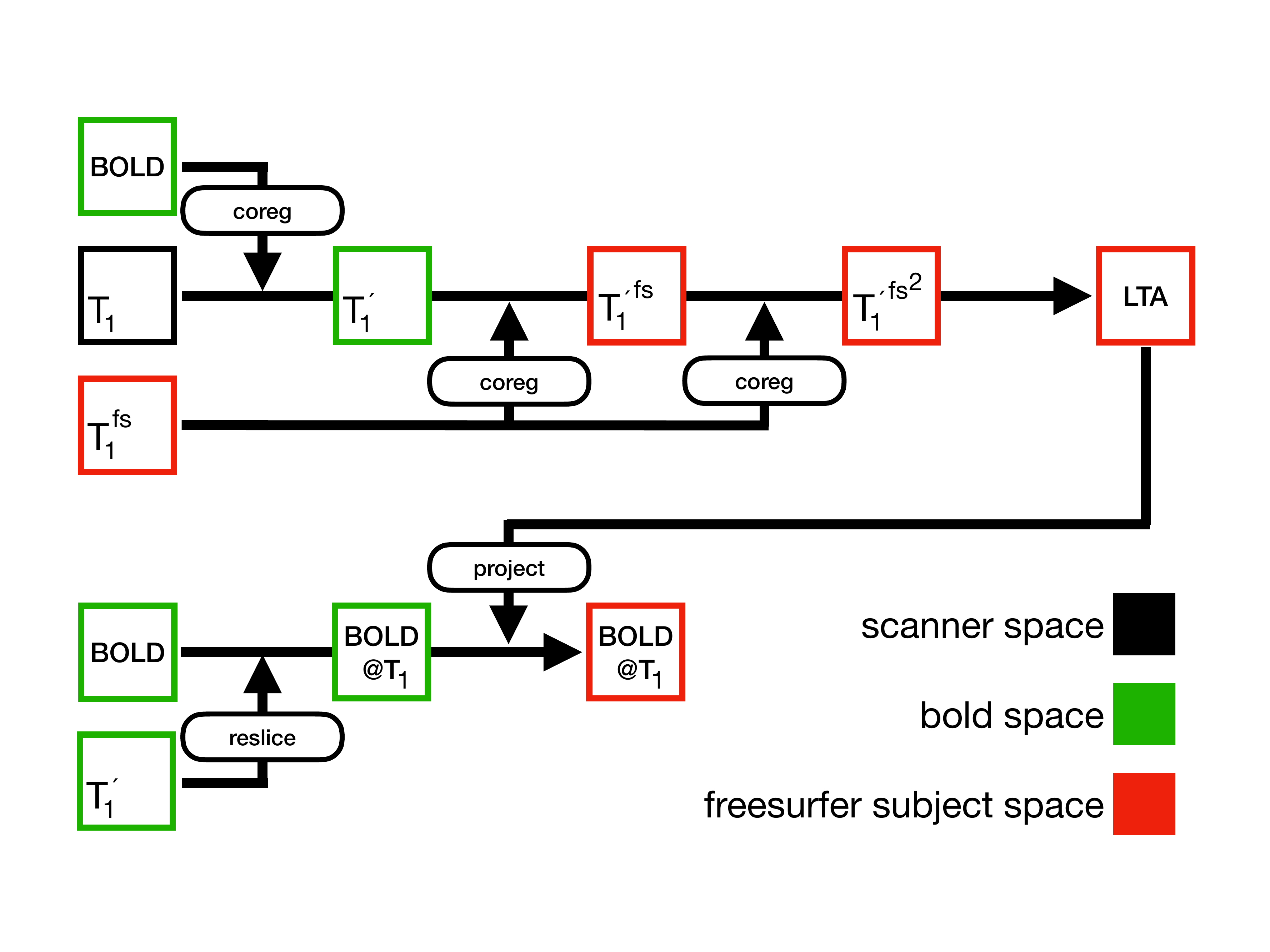}
    \caption{ \small Volume-to-surface processing pipeline.}
    \label{fig:processing}
\end{figure}
To reduce computational load, both the left and right pial surface models were resampled to 10,000 vertices per hemisphere. This was accomplished using the Workbench command \texttt{-surface-resample}, which leverages the registration between a sphere consisting of 10,000 vertices (created using the Workbench command \texttt{-surface-create-sphere}) and the participant's spherical surface generated using FreeSurfer. The BOLD data was then projected to the left and right hemisphere resampled surfaces using the Workbench command \texttt{-metric-resample}. Finally, the FreeSurfer labeling of four sensorimotor areas (i.e., the paracentral gyrus, postcentral gyrus, precentral gyrus, and caudal middle frontal gyrus \citep{Verstraete:2011fp}; all were taken from the Desikan-Killiany atlas\citep{desikan-2006-autom-label}) were resampled to 10,000 vertices using the Workbench command \texttt{-label-resample}. These labels were combined to produce a participant-specific motor mask to limit the location of statistical estimation. 

Before model fitting, we identified and removed noisy volumes based on data-driven leverage scrubbing using the \texttt{fMRIscrub} R package \citep{mejia2017pca} (version 0.1.2), which identifies volumes that differ substantially from the multivariate distribution of images. We employed a threshold of 4 times the median leverage for scrubbing.  We also excluded any sessions where more than 25\% of volumes were scrubbed. This resulted in exclusion of one visit from one ALS participant, one visit each from three HC participants, and two visits from one HC participant. 

 \begin{figure}
    \centering
    \includegraphics[width=5in]{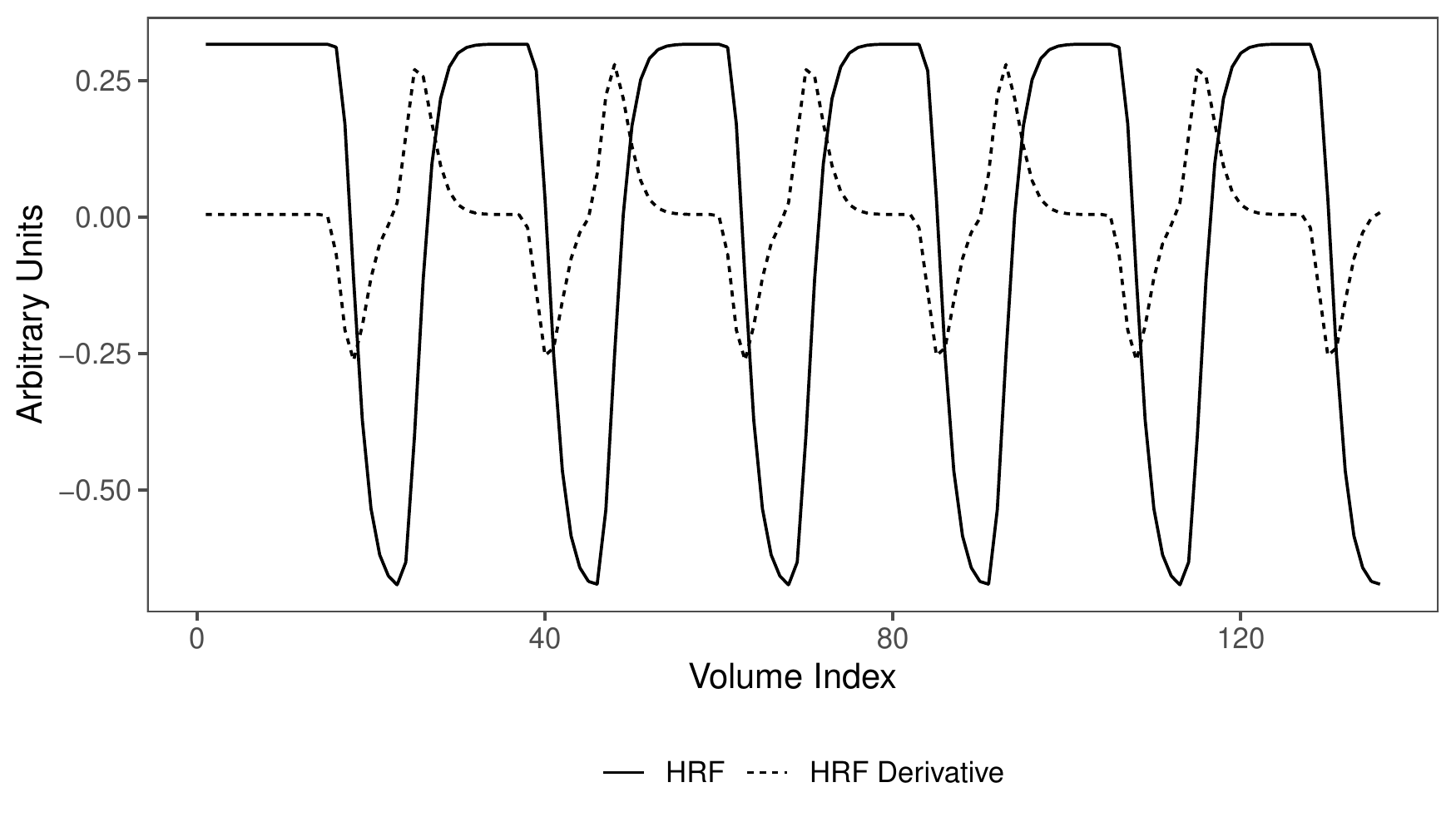}
    \caption{ \small Modeled hemodynamic response (HRF) for the right hand clench task and its temporal derivative (dHRF).}
    \label{fig:HRF}
\end{figure}

In model (\ref{eqn:lmer_ALS}), the function $f(\cdot)$ is a natural cubic spline, which allows for a non-linear relationship between hand motor disability and activation size. Spline knots were placed at the 33rd and 67th quantiles. Natural splines have boundary conditions that enforce a linear fit beyond the boundary knots, which avoids the extreme boundary fits often observed in standard polynomial regression.  A basis was generated in \texttt{R} using the \texttt{ns} function from the \texttt{splines} package version 4.0.3. 

This final model form in equation (\ref{eqn:lmer_ALS}) was determined by a series of likelihood ratio tests.  We compared three models for each predictor: one assuming a linear fit, one allowing a non-linear spline fit, and one excluding the predictor.  Each test was based on left-hemispheric (contra-lateral) activation at an effect size of $\gamma=0\%$, which provided the most robust areas of activation. We found statistically significant evidence for a non-linear relationship between hand disability and size of activation and a linear relationship between other disability and size of activation. We also considered an alternative model with days since onset as a predictor (spline or linear fit) in place of the disease burden measures. These models were substantially worse in terms of predictive accuracy and Akaike information criterion (AIC) \citep{akaike1998information}.

\newpage
\section{Longitudinal spatial Bayesian task fMRI analysis}
\label{app:model}

\begin{figure}[H]
    \centering
    \includegraphics[width=4in]{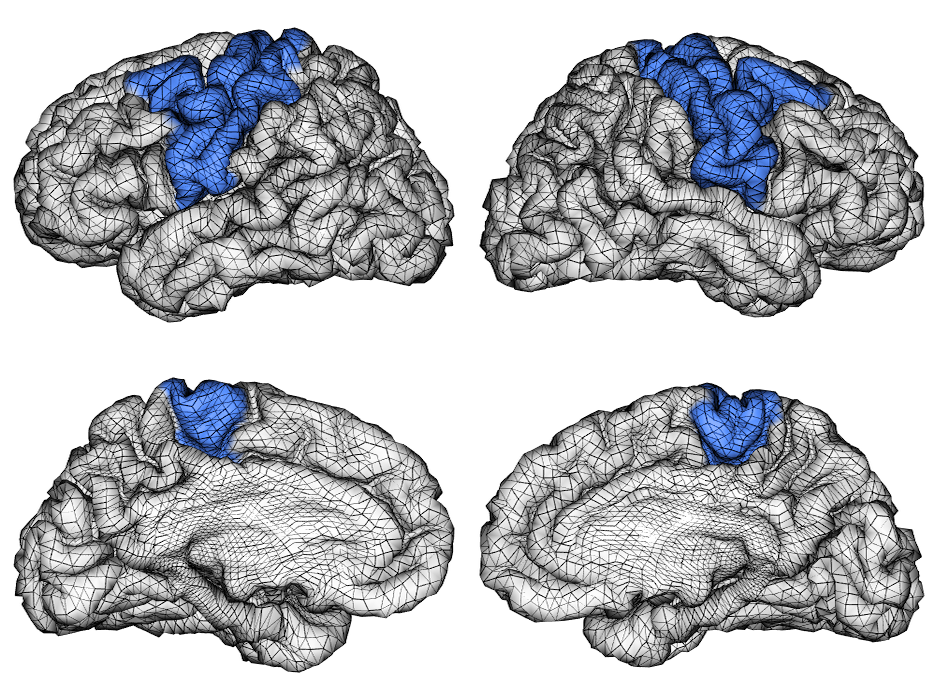}
    \caption{ \small Triangular mesh for the resampled pial surface of each hemisphere for one individual with ALS. The model is fit within the motor cortex, which is shaded in blue.  For this individual, the motor cortex includes 1,455 resampled vertices in the left hemisphere and 1,524 in the right hemisphere.}
    \label{fig:mesh_ALS}
\end{figure}

The model is fit within each hemisphere on the brain separately. The triangular mesh representing the participant-specific cortical surface, after resampling and masking as described above, contained approximately 1,500 vertices per hemisphere.  The exact size and shape varied across participants due to differences in cortical anatomy.  Fig.\ \ref{fig:mesh_ALS} shows the surface meshes for one participant with ALS.  

In our longitudinal spatial Bayesian modeling framework, the amplitudes and areas of activation are estimated for each visit, but model parameters including the residual variance and spatial properties of the task activation fields (e.g. correlation range, variance) are estimated using data from multiple visits to improve estimation efficiency. The same set of tasks must be performed across visits, though the stimulus timing can vary over visits.

Consider a single subject and hemisphere.  Let $j=1,\dots,J$ index visits and $k=1,\dots, K$ index task stimuli. In our models, $K=2$ (the canonical HRF and its first derivative), and the number of visits per participant varied between $J=3$ to $J=10$. Let $T_j$ be the number of volumes in visit $j$ after scrubbing, and let $V$ be the number of surface vertices within the mask. Note that the surfaces are required to be spatially co-registered across visits within a subject, but not across subjects, as the model is fit separately for each subject. Let $\bfy_j(v)$ ($T_j\times 1$) be the processed and scrubbed fMRI data at vertex $v$. Let $\bfx_{jk}$ ($T_j\times 1$) represent the expected BOLD response to task $k$ (excluding scrubbed volumes).
In the classical GLM, we would fit a separate linear model at each location $v=1,\dots,V$, namely
\begin{equation}\label{eqn:classical_GLM}
\bfy_j(v) = \sum_{k=1}^K \bfx_{jk}\beta_{jk}(v) + \bfepsilon_j(v), 
\quad \bfepsilon_j(v)\sim N(\bfzero, \sigma^2\bfI_{T_j}),
\end{equation}

where $\beta_{jk}(v)$ is the activation amplitude associated with task $k$.  In equation (\ref{eqn:classical_GLM}) the residuals are assumed to be temporally independent, which can be achieved by prewhitening.

To illustrate the construction of our longitudinal spatial Bayesian GLM, we first combine across vertices within a single session, describe the incorporation of spatial priors on the task amplitudes to yield a spatial Bayesian model, then generalize to the longitudinal case.  Denote

\begin{equation}
\bfy_j = \begin{bmatrix}\bfy_j(1) \\ \vdots \\ \bfy_j(V)\end{bmatrix},\quad
 \bfX_{jk} = \bfI\otimes{\bfx}_{jk} = \begin{bmatrix} \bfx_{jk} & & \\ & \ddots & \\ & & \bfx_{jk} \end{bmatrix} ,\quad
  \bfbeta_{jk} = \begin{bmatrix}\beta_{jk}(1) \\ \vdots \\ \beta_{jk}(V)\end{bmatrix}, \text{ and }
\bfepsilon_j = \begin{bmatrix}\bfepsilon_j(1) \\ \vdots \\ \bfepsilon_j(V)\end{bmatrix},
\end{equation}

where $\otimes$ denotes the Kronecker product.  Then we can write the single-session model as

\begin{equation}
\bfy_j = \bfX_{jk} \bfbeta_{jk} + \bfepsilon_j, \quad \bfepsilon_j\sim N(\bfzero, \sigma^2\bfI).
\end{equation}

Assuming spatial process priors on the $\bfbeta_{jk}$, $k=1,\dots, K$, along with hyperpriors on their parameters, yields a spatial Bayesian model.  \cite{mejia2020bayesian} proposed employing a class of flexible Gaussian Markov random field (GMRF) priors that are appropriate for high-dimensional data in a triangular mesh format, known as stochastic partial differential equation (SPDE) priors \citep{lindgren2011spde}. Specifically, SPDE priors are zero-mean multivariate Normal priors with a sparse precision (inverse covariance) structure. The precision matrix has non-zero entries along the diagonal and in cells corresponding to neighboring locations in the triangular mesh.  We provide more details on the precision structure in the specification of the longitudinal model below.  

Now combining over sessions, denote

\begin{equation}
\bfy \begin{bmatrix}\bfy_1 \\ \vdots \\ \bfy_J\end{bmatrix},\quad
\bfX_k = \begin{bmatrix} \bfX_{i1k} & & \\ & \ddots & \\ & & \bfX_{iJk} \end{bmatrix},\quad
\bfbeta_k = \begin{bmatrix}\bfbeta_{1k} \\ \vdots \\ \bfbeta_{Jk}\end{bmatrix}, \text{ and }
\bfepsilon = \begin{bmatrix}\bfepsilon_1 \\ \vdots \\ \bfepsilon_J\end{bmatrix}.
\end{equation}

The longitudinal spatial Bayesian model is given by
\begin{align}\label{eqn:Bayes_GLM}
\begin{split}
(\bfy|\bfbeta_1,\dots,\bfbeta_K)
&= \sum_{k=1}^K \bfX_k\bfbeta_k +\bfepsilon \\
\bfepsilon|\sigma^2 &\sim N\left(\bfzero,\sigma^2\bfI\right) \\
\bfbeta_{jk}|\kappa_k,\tau_k &\stackrel{iid}{\sim} N(\bfzero, \bfQ_k^{-1})\text{ for }j=1,\dots,J,\ k=1,\dots,K \\
\bftheta &\sim \pi(\bftheta),
\end{split}
\end{align}

where $\bftheta=(\kappa_{1},\tau_{1},\dots,\kappa_K,\tau_K,\sigma^2)$ are all of the hyperparameters and $\pi(\bftheta)$ is their joint prior density.  We assume independent log-normal priors on the spatial hyperparameters $\kappa_k$ and $\tau_k$ and a gamma prior on the inverse residual variance. Note that the spatial hyperparameters are allowed to vary across tasks, allowing for differences in the spatial properties of different tasks, but are common across visits, which improves estimation efficiency.  The form of the spatial precision with parameters $\kappa$ and $\tau$ is $\bfQ = \tau(\kappa^4\bfC + 2\kappa^2\bfG + \bfG\bfC^{-1}\bfG)$, where $\bfC$ is a diagonal matrix and $\bfG$ is a sparse symmetric matrix with non-zero entries in cells corresponding to neighboring vertices in the triangular mesh \citep{lindgren2015bayesian}.  The parameter $\kappa$ controls the spatial dependence of the field, while $\tau$ controls its variance.  

This model can be estimated using the \texttt{BayesfMRI} R package, which uses R-INLA \citep{lindgren2015bayesian} to compute the necessary posterior quantities for each latent field $\bfbeta_{jk}$, as described in detail in \cite{mejia2020bayesian}. Given the posterior mean and precision of each latent field, we can then identify areas of activation based on the joint posterior distribution using an excursions set approach \citep{bolin2015excursion, mejia2020bayesian}.  This avoids massive multiple comparisons and results in much greater power to detect true activations by leveraging spatial dependencies and avoiding multiplicity correction. 

Areas of activation can also be identified through \texttt{BayesfMRI}, which uses the \texttt{excursions} package \citep{bolin2018excursions} to identify areas exceeding a specified effect size $\gamma$ (e.g. 1\% signal change) at a given significance level $\alpha$. For more information on the model estimation and computation of excursions sets, see \cite{mejia2020bayesian}.

\newpage
\section{Computation Time}\label{app:computation}

\begin{figure}[H]
    \centering
    \includegraphics[page=1, width=3in]{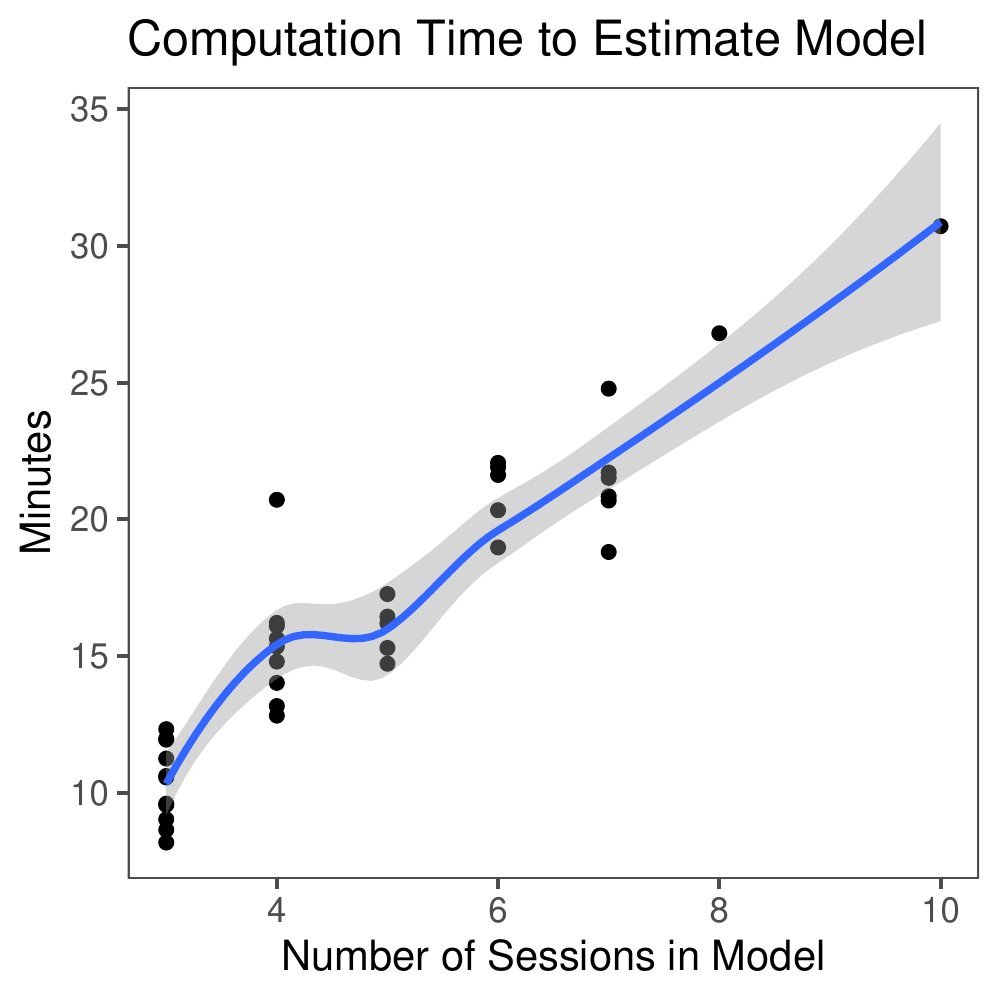}
    \includegraphics[page=2, width=3in]{plots/comptime.pdf}
        \caption{ \small Computation time in minutes for model estimation and identifying areas of activation, per participant. Times represent the sum across both hemispheres per participant. Time to identify activations reflects the average time per visit, averaged over visits, at a given effect size. Computation times for both model estimation and identifying areas of activation grew approximately linearly with the number of visits per participant. }
    \label{fig:comptime}
\end{figure}

All computations were performed in R version 4.0.3 \citep{Rproject} using the \texttt{BayesfMRI} package on a Mac Pro computer with a 2.7 GHz 24-core Intel Xeon W processor with 512 GB of memory. Depending on the number of visits being simultaneously estimated, model estimation per participant and hemisphere took 10 to 30 minutes and required approximately 10 to 25 GB of RAM.  Identifying areas of activation took an additional 1-5 minutes per session and effect size. Computation times for all participants are shown in Fig.\ \ref{fig:comptime}. 

\newpage
\section{Validation of Spatial Bayesian GLM}
\label{app:validation}

\begin{figure}
\centering
\begin{tabular}{ccc}
& {\large Activation Amplitude} & {\large Area of Activation} \\[5pt]
\begin{picture}(0,60)\put(-5,30){\rotatebox[origin=c]{90}{Bayesian GLM}}\end{picture} & 
\fbox{\includegraphics[width=2.8in, trim=0 15cm 0 3cm, clip]{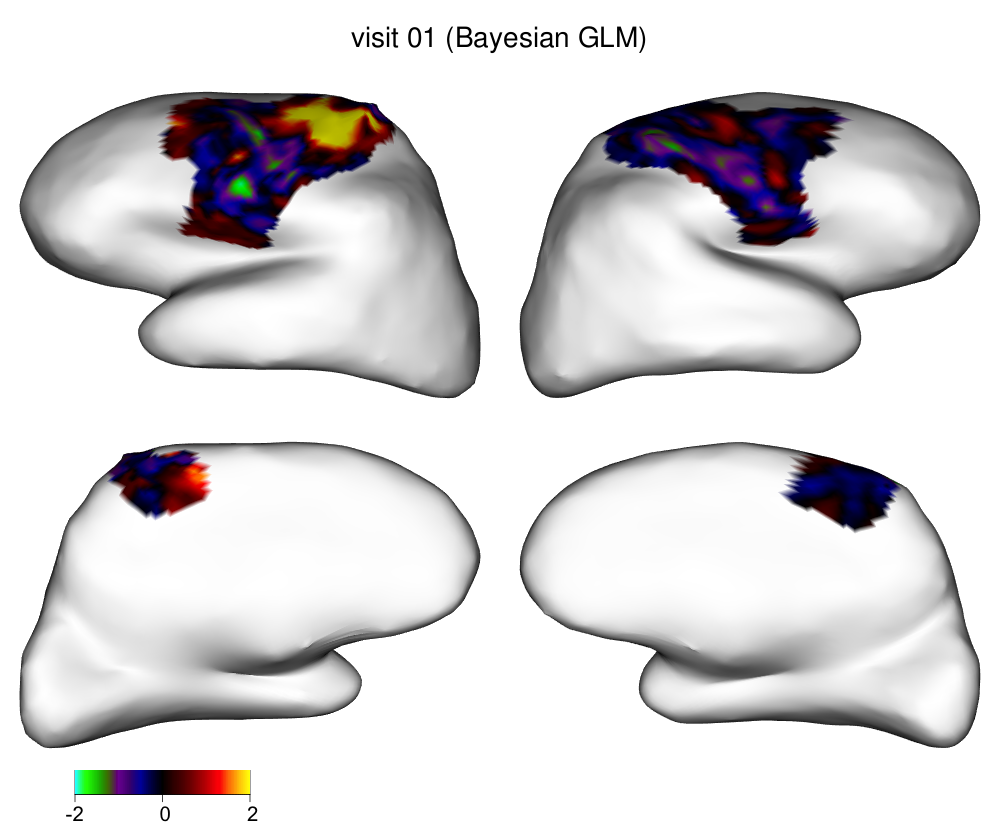}} &
\fbox{\includegraphics[width=2.8in, trim=0 12.5cm 0 3cm, clip]{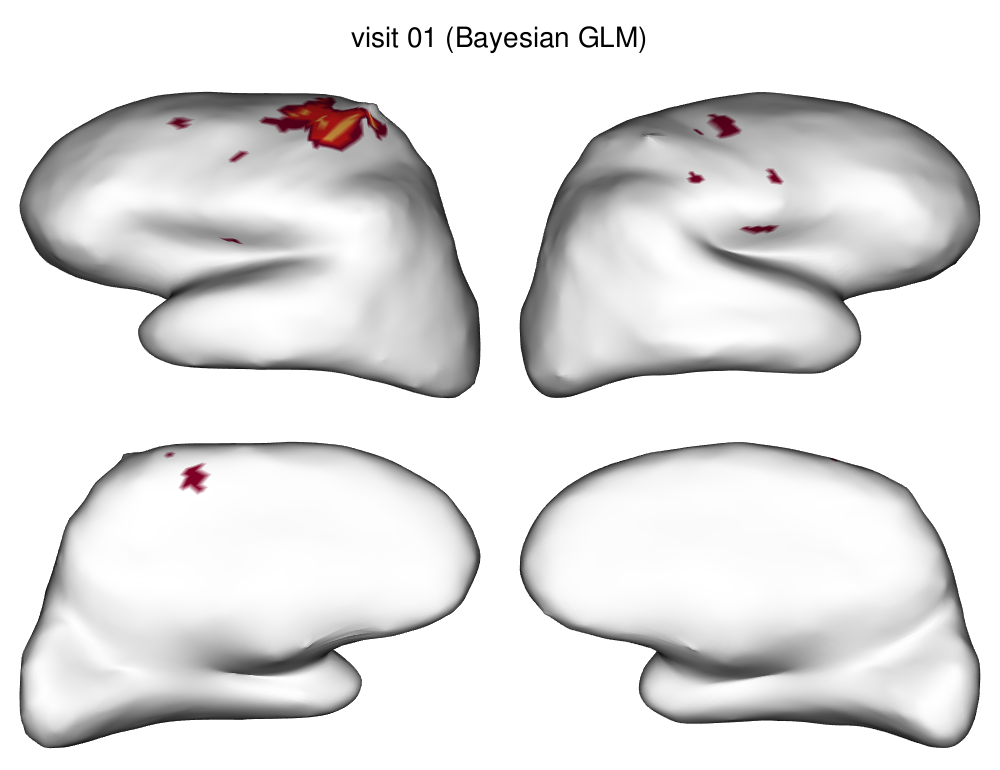}} \\[5pt]
& \includegraphics[width=1.3in]{images/legend_zlim2.png} & \actlegendbay \\[5pt]
\begin{picture}(0,60)\put(-5,30){\rotatebox[origin=c]{90}{Classical GLM}}\end{picture} & 
\fbox{\includegraphics[width=2.8in, trim=0 15cm 0 3cm, clip]{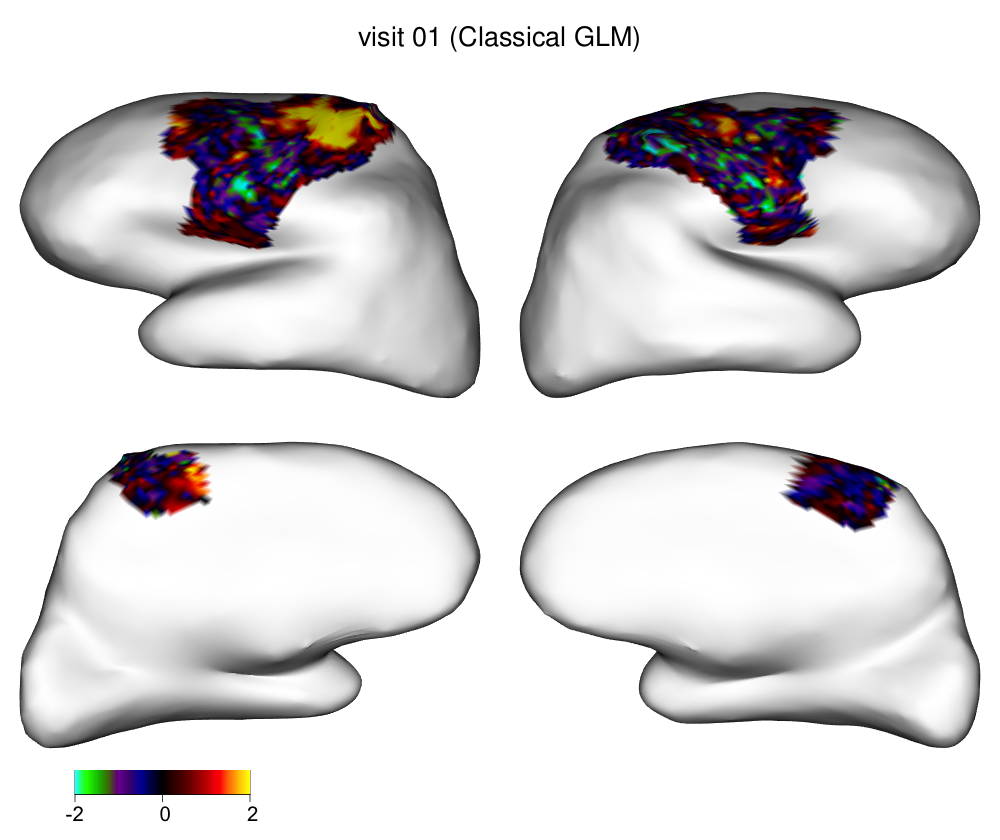}} &
\fbox{\includegraphics[width=2.8in, trim=0 12.5cm 0 3cm, clip]{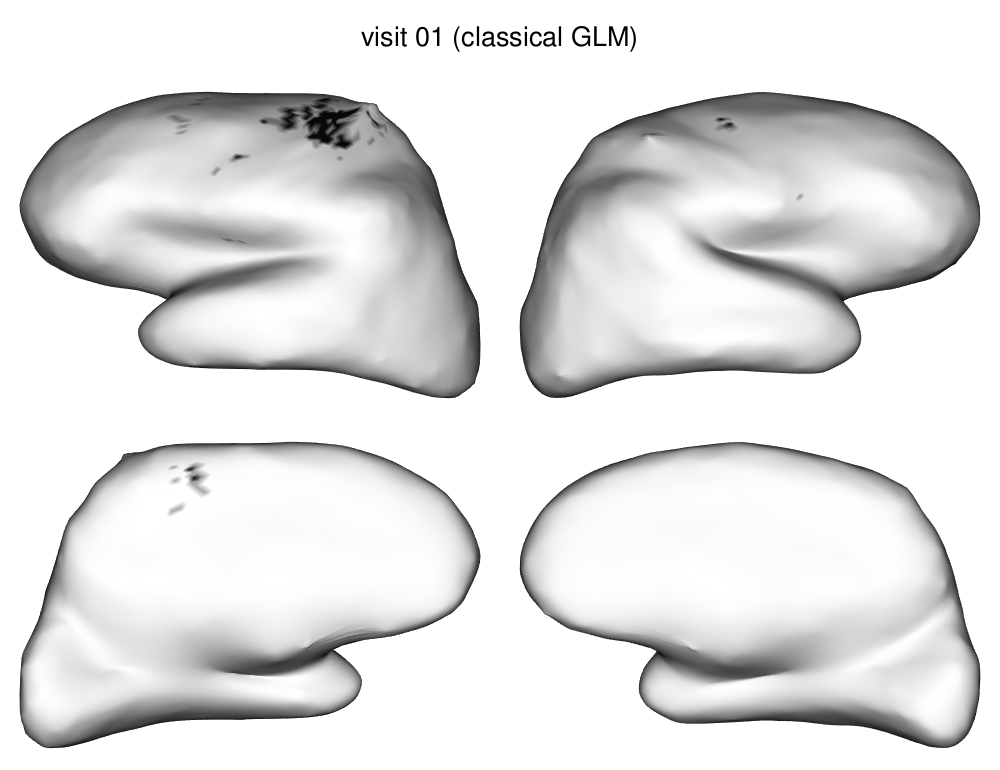}} \\[5pt]
& \includegraphics[width=1.3in]{images/legend_zlim2.png} & \actlegendclass \\[5pt]
\end{tabular}
\caption{\small \textbf{Bayesian GLM and classical GLM estimates of activation amplitude and areas of activation in one HC participant.} The Bayesian GLM tends to produce estimates that are smoother and areas of activation that are larger and more contiguous.}
\label{fig:estimates_classical}
\end{figure}

We first compared the results of the Bayesian and classical approaches visually for one example HC participant in Fig.\ \ref{fig:estimates_classical}.  The left panel shows estimates of activation amplitude produced from each approach. The Bayesian GLM produced amplitudes of activation that were noticeably smoother than those produced by the classical GLM. This is due to the implicit smoothing in the model estimation for the spatial Bayesian GLM, which accounts for spatial dependencies between neighboring vertices.  The degree of smoothing is determined in an optimal fashion and avoids smoothing of noise along with the signal as in data smoothing \citep{lindquist2015zen}.   

The right panel of Fig.\ \ref{fig:estimates_classical} shows areas of activation produced from both approaches.  Note that the Bayesian GLM with effect size $\gamma=0\%$ is analogous to the classical GLM with FWER correction, since both provide similar guarantees around false positive control, and setting $\gamma=0\%$ is comparable to the null hypothesis of no activation.  Yet the Bayesian GLM produced much larger areas of activation at $\gamma=0\%$ compared to the classical GLM with FWER correction. This is due to the power gained in the Bayesian GLM by leveraging spatial dependencies and avoiding the need for multiplicity correction. FDR correction produced larger areas of activation than FWER correction, but in this participant they were still smaller than those produced with the Bayesian GLM at $\gamma=0\%$, and do not provide similar guarantees around false positive control. 

To quantitatively assess the quality of the areas of activation produced by the Bayesian GLM and classical GLM using different methods and effect sizes, we analyzed the longitudinal stability of the size of those areas in HC participants. Since we do not expect much change in HC participants over the duration of the study, smaller variation in the size of activation over time was considered better.

To quantitatively assess the quality of the areas of activation produced by the Bayesian GLM and classical GLM using different methods and effect sizes, we analyzed the longitudinal stability of the size of those areas in HC participants. Since we do not expect much change in HC participants over the duration of the study, smaller variation in the size of activation over time was considered better. We compared the Bayesian GLM at the three effect sizes ($\gamma=0\%$, $1\%$, $2\%$) and the classical GLM using FWER and FDR correction.  Since some of these methods tend to result in larger areas of activations, they will tend to have larger variance (since variance is not unit-less), so it is important to consider the size of activation when comparing the variance. 

\begin{figure}
    \centering
    \begin{subfigure}[b]{0.48\textwidth}
    \includegraphics[width=3in, page=2, trim=0 0 0 7mm, clip]{plots/variance_HCs.pdf}
    \caption{ \small SD versus mean activation size within HCs}
    \end{subfigure}    
    \begin{subfigure}[b]{0.48\textwidth}
    \includegraphics[width=3in, page=1, trim=0 0 0 7mm, clip]{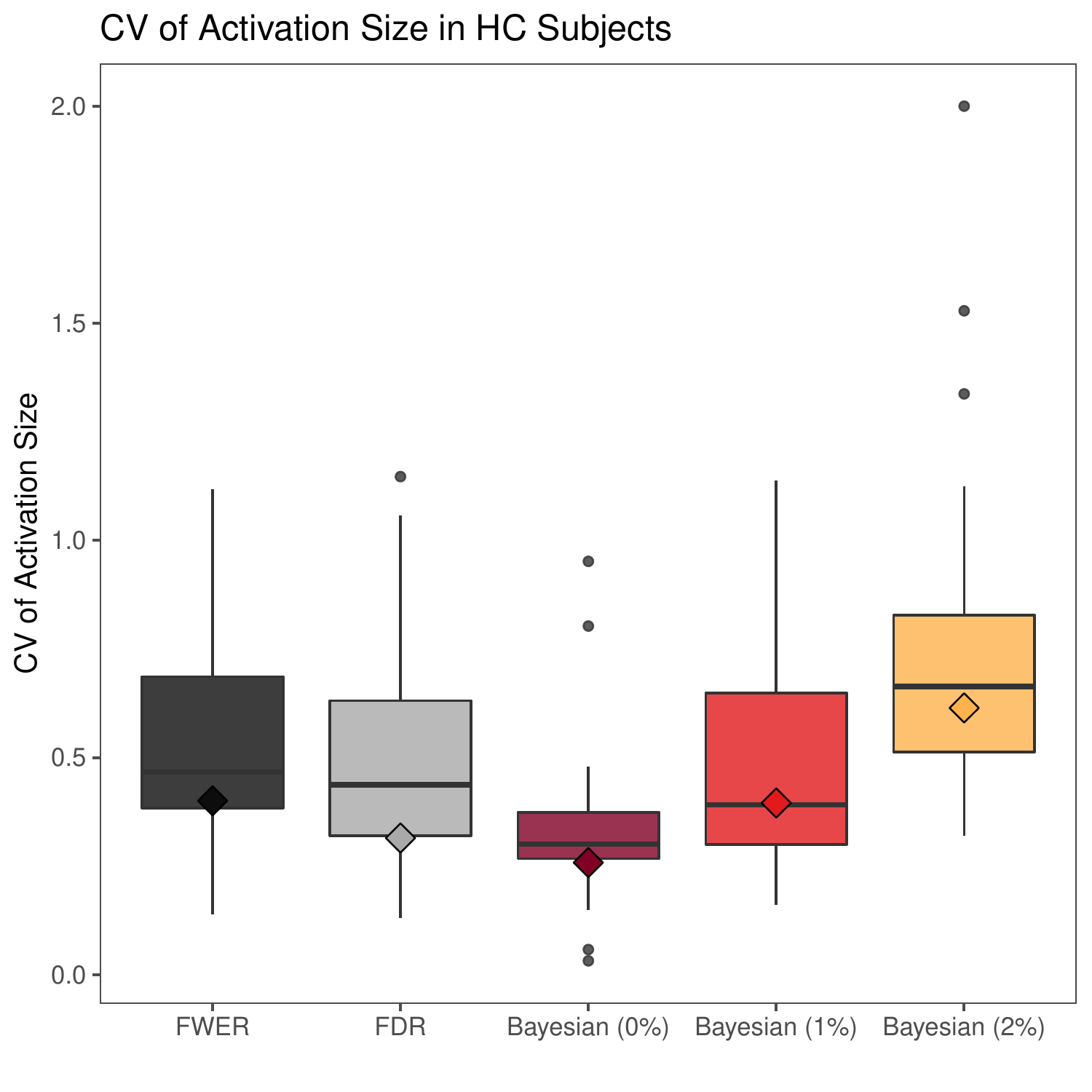}
    \caption{ \small CV of activation size within HCs}
    \end{subfigure}
    \caption{ \small \textbf{Longitudinal reliability of size of activations in HC participants.} \textbf{(a)} Standard deviation (SD) of size of activations versus mean size of activations across visits. Each point represents a HC participant and a method. Lines represent a linear regression fit for each method. The Bayesian GLM results in lower variation in size of activation across visits compared with the classical GLM, considering the mean size of activation. For example, when Bayesian GLM with $\gamma=0\%$ and FDR correction result in similar mean sizes of activation, the Bayesian GLM has lower within-participant variance. Similarly, when Bayesian GLM with $\gamma=1\%$ and FWER correction produce activations of similar size, the Bayesian GLM has lower within-participant variance.  \textbf{(b)} Coefficient of variation (CV) in activation size. For each method and effect size, boxplots display the longitudinal CV \textit{within} each HC participant; the colored diamonds display the CV \textit{between} HC participants, based on the mean across visits for each participant. Methods with lower within-participant CV produce more reliable estimates, and those with higher between-participant CV better respect differences between participants. Therefore, the best case scenario is \textit{low} within-participant CV (boxplot) and \textit{high} between-participant CV (diamond). The best methods in this respect are the Bayesian GLM at $\gamma=0\%$ and $\gamma=1\%$.}
    \label{fig:var_HC}
\end{figure}

Fig.\ \ref{fig:var_HC} displays two plots of longitudinal variation in size of activation within HC participants. Both plots illustrate that the Bayesian GLM results in lower variation in size of activation across visits compared with the classical GLM, considering size of activation. In Fig.\ \ref{fig:var_HC}(a), we plot the standard deviation (SD) across visits versus the mean across visits. The line from a linear model relating the SD to the mean for each method is also shown. We observe that, considering mean size of activation, the Bayesian GLM results in lower variation in size of activation across visits compared with the classical GLM. For example, Bayesian GLM with an effect size of $\gamma=0\%$ and classical GLM with FDR correction often result in similar mean sizes of activation, but the Bayesian GLM has lower variance within HC participants over time. Similarly, the Bayesian GLM with $\gamma=1\%$ and classical GLM with FWER correction often produce activations of similar size, but the Bayesian GLM has lower within-participant variance. 

In Fig.\ \ref{fig:var_HC}(b) we explicitly account for differences in the size of activation through the coefficient of variation (CV), a unit-less measure of variability equal to the standard deviation divided by the mean. For each method and effect size, boxplots display the longitudinal CV \textit{within} each HC participant.  Lower within-participant CV indicates more reliable estimates. This plot shows that the Bayesian GLM with $\gamma=0\%$ produces highly reliable areas of activation in HC participants. The colored diamonds display the CV \textit{between} HC participants, based on the mean across visits for each participant. Methods  that produce higher between-participant CV better preserve differences between participants. The Bayesian GLM with $\gamma=1\%$ and $\gamma=2\%$ perform the strongest in this regard, since they have higher between-participant CV relative to the within-subject CV.
 
\newpage
\section{Additional Results Figures}
\label{app:more_result_figures}

\begin{figure}[H]
    \begin{subfigure}[b]{1\textwidth}
    \centering
    \includegraphics[height=2.7in, page=1, trim=0 1.5cm 0 0, clip]{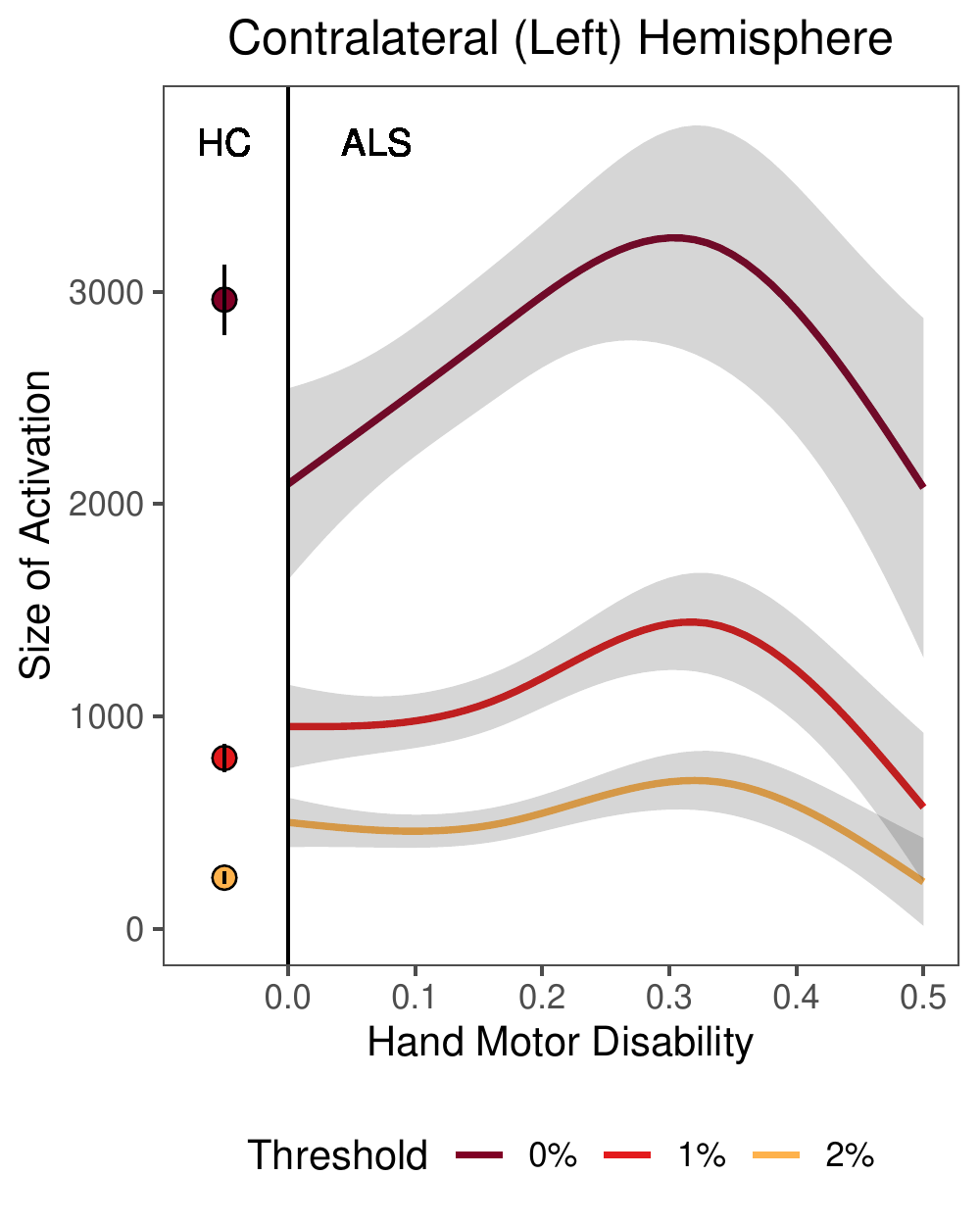}
    \includegraphics[height=2.7in, page=2, trim=6mm 1.5cm 0 0, clip]{plots/lmer_activation_vs_diability_lh.pdf}   
    \end{subfigure}
    \begin{subfigure}[b]{1\textwidth}
    \centering
    \vspace{5mm}
    \includegraphics[height=2.7in, page=1, trim=0 1.5cm 0 0, clip]{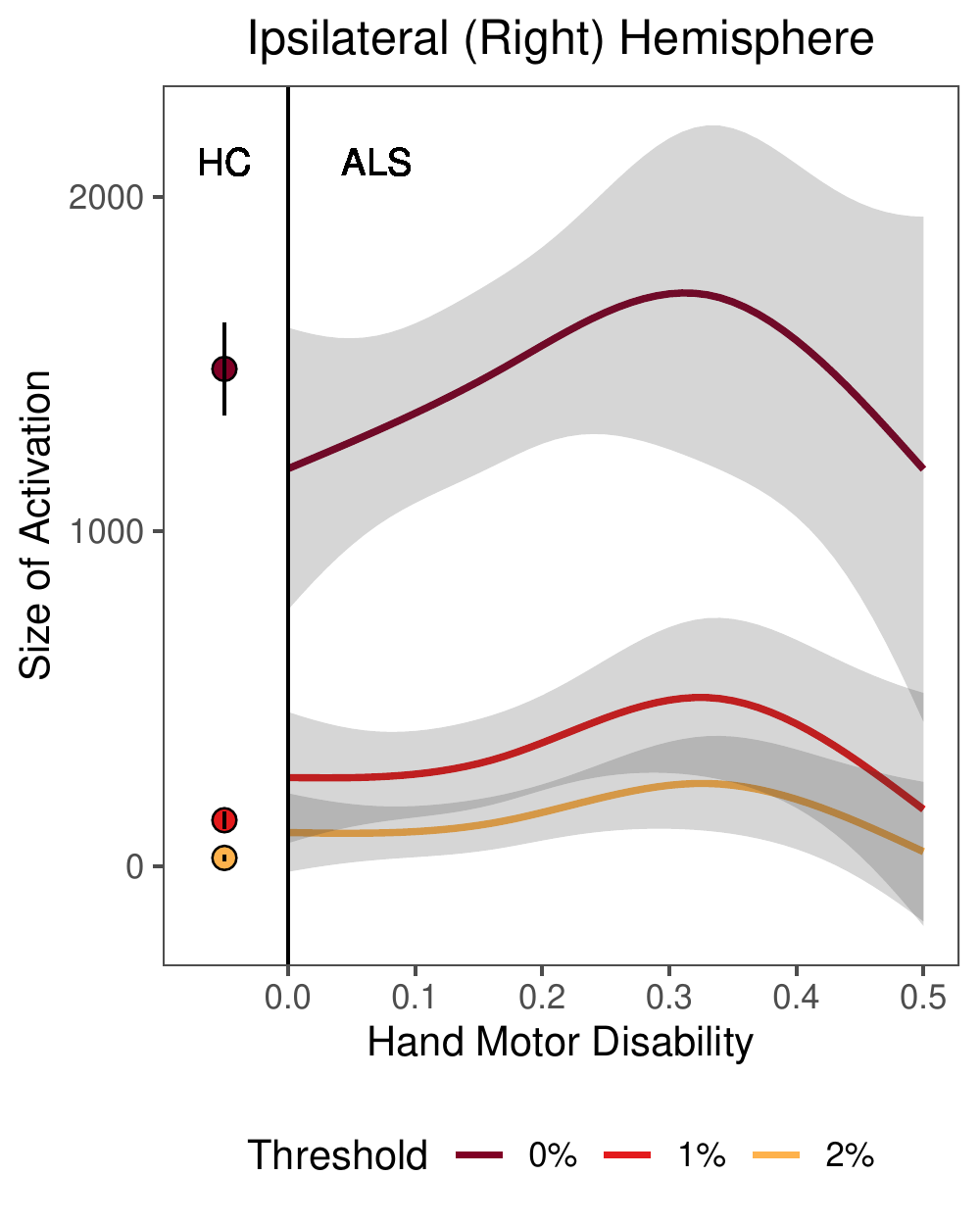}
    \includegraphics[height=2.7in, page=2, trim=6mm 1.5cm 0 0, clip]{plots/lmer_activation_vs_diability_rh.pdf} \\
     \includegraphics[width=2in]{plots/lmer_activation_vs_diability_legend.pdf}
    \end{subfigure}
        \caption{ \small Coefficient curves for the size of the area of activation in response to right hand clenching, based on the mixed effects model given in equation (\ref{eqn:lmer_ALS}), including subject A04. The curves are very similar to those observed in the main text Fig.\  \ref{fig:mixed_effects_model}. This illustrates that the relationships observed between ALS disability and size of activation are robust to the inclusion or exclusion of this subject.}
    \label{fig:mixed_effects_model_A04}
\end{figure}

One individual with ALS (A04) had a very slow disease trajectory (see Fig.\ \ref{fig:ALSFRS}) and had many more visits spanning a much longer duration, compared with  other ALS participants (Fig.\ \ref{fig:visit_timing}). To avoid undue influence of this unusual individual on the random intercept models given in equation (\ref{eqn:lmer_ALS}), participant A04 was excluded from model fitting. Here, we present the results of the models for each hemisphere and effect size including this individual.  Fig.\ \ref{fig:mixed_effects_model_A04} shows coefficient curves that are very similar to those seen in the main text. This illustrates that the relationships observed between ALS disability and size of activation are robust to the inclusion or exclusion of this individual.

\begin{figure}[H]
\centering
    \includegraphics[height=4in]{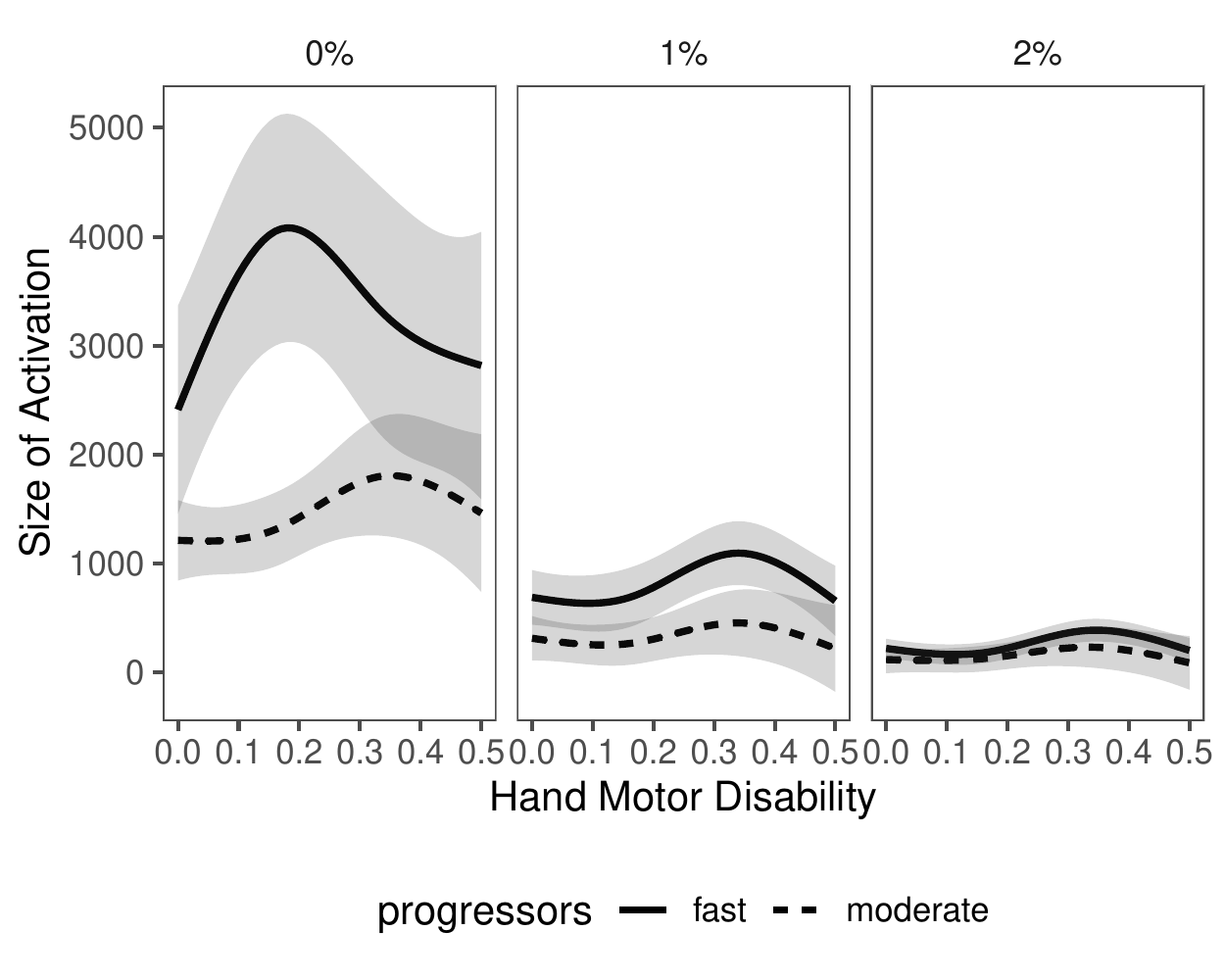}
        \caption{ \small Coefficient curves for the size of \textit{ipsilateral} activation in response to right hand clenching, based on the mixed effects model given in equation (\ref{eqn:lmer_ALS}), by progression rate.  Here we focus on the relationship between activation size and Hand Motor Disability, with Other Disability fixed at zero. The curves are very similar to those observed in the main text Fig.\ \ref{fig:mixed_effects_model_fastslow}.}
    \label{fig:mixed_effects_model_fastslow_rh}
\end{figure}

\begin{figure}[H]
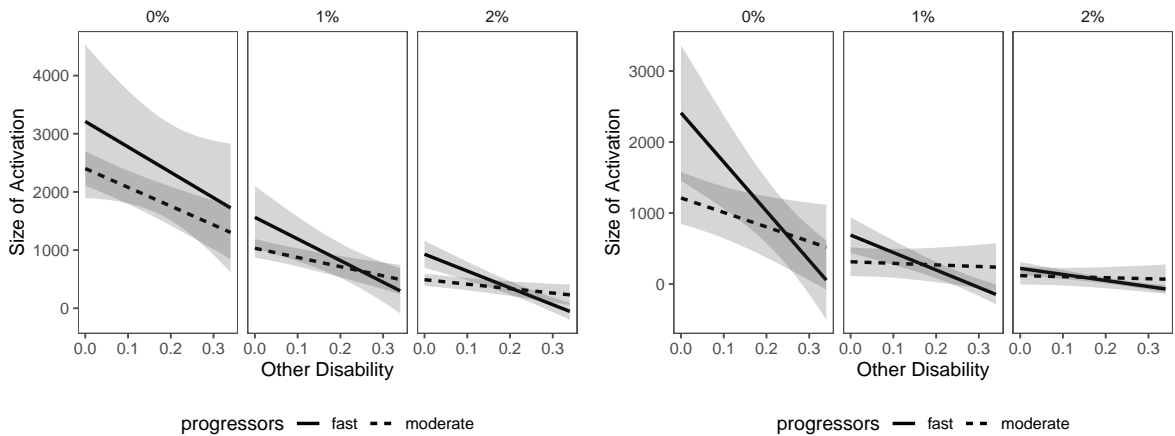

\begin{subfigure}[b]{0.485\textwidth}
    \includegraphics[width=1\textwidth, page=2]{plots/lmer_activation_vs_diability_lh_fastslow.pdf}
    \caption{Contralateral (Left Hemisphere) Activation}
\end{subfigure}
\begin{subfigure}[b]{0.485\textwidth}
    \includegraphics[width=1\textwidth, page=2]{plots/lmer_activation_vs_diability_rh_fastslow.pdf}
    \caption{Ipsilateral (Right Hemisphere) Activation}
\end{subfigure}        
\caption{ \small Coefficient curves for the size of activation in response to right hand clenching, based on the mixed effects model given in equation (\ref{eqn:lmer_ALS}), by progression rate.  Here we focus on the relationship between activation size and Other Disability, with Hand Motor Disability fixed at zero. Fast progressors show larger size of activation at low levels of disability and decline faster as a function of disability, compared with moderate progressors.}
    \label{fig:mixed_effects_model_fastslow_other}
\end{figure}

\end{document}